\documentclass[twocolumn]{aastex701}

\usepackage{amsmath}
\usepackage{enumitem}
\usepackage{newtxtext,newtxmath}
\usepackage{bm}
\usepackage{graphicx}
\usepackage{subcaption}
\usepackage{booktabs}
\usepackage{multirow}
\usepackage{makecell}

\newcommand{\Msun}{\,{\mathrm{M}_\odot}}

\newcommand{\feh}{\,{\mathrm{[Fe/H]}}}

\shorttitle{Globular Cluster Systems of BCGs}
\shortauthors{Smith et al.}
\submitjournal{OJAp}

\begin{document}

\title{Modeling Globular Cluster Systems of the Giant Galaxies: Evolution from Redshift 1}

\author[0009-0008-8293-9404]{Isaac Smith}
\affiliation{Department of Astronomy, University of Michigan, Ann Arbor, MI 48109, USA}
\email[show]{isaacsmi@umich.edu}  

\author[orcid=0000-0001-9852-9954]{Oleg Y. Gnedin}
\affiliation{Department of Astronomy, University of Michigan, Ann Arbor, MI 48109, USA}
\email[]{ognedin@umich.edu}

\author[0000-0002-5970-2563]{Yingtian Chen}
\affiliation{Center for Cosmology and AstroParticle Physics (CCAPP), The Ohio State University, Columbus, OH 43210, USA}
\affiliation{Department of Astronomy, The Ohio State University, Columbus, OH 43210, USA}
\email{chen.15712@osu.edu}

\begin{abstract}
Globular cluster (GC) systems provide unique probes of the evolution of their host galaxies. James Webb Space Telescope (JWST) imaging in high-redshift galaxy clusters has extended the studies of GC populations over 7~Gyr of cosmic history but detects only the most massive GCs. To complement these observations, we present theoretical predictions for the full populations of GCs around distant brightest cluster galaxies (BCGs). The model has been validated on the observed mass, metallicity, spatial and kinematic distributions of GC systems in Milky Way-sized galaxies. Here we apply it to giant BCG systems. The GC mass function is expected to become slightly shallower with time as low-mass GCs are preferentially disrupted, but this evolution is challenging to detect in current observational samples limited to very massive GCs. Outside the BCG core, the model surface density profiles are generally consistent with a power law of projected distance from the galaxy center, with a wide range of slopes $\gamma\approx 1.0-1.8$ consistent with observations. Accretion of massive satellite galaxies at $z<1$ can dominate the overall GC system, and major mergers can lead to a significant spatial redistribution of in-situ GCs. The model predicts a mass-metallicity relation for GCs above $10^6\Msun$, similar to the blue tilt observed in massive galaxies in the local universe. Because of the tilt, the observed samples in high-redshift systems are expected to be more metal-rich and represent a higher in-situ fraction relative to the full GC population. The crossover radius between the in-situ and ex-situ GC populations can vary significantly between host galaxies of similar masses and is a sensitive probe of the assembly history of the host galaxy.
\end{abstract}


\section{Introduction} 
\label{sec:intro} 

Globular clusters (GCs) are old stellar systems which populate the halos of massive galaxies. They are thought to trace the assembly history of their host galaxies. A precise understanding of globular cluster formation and evolution is necessary to uncover their role in galaxy formation. We rely on theoretical models of GC formation and dynamical disruption, which must be validated with observations.

Until recently, observing globular clusters at redshifts greater than $z \approx 0.04$ was extremely difficult. Most distant samples included the brightest cluster galaxies (BCGs) imaged with the Hubble Space Telescope \citep{ Harris2014, Harris2016, Harris2017a, Harris2017b}. Now, the James Webb Space Telescope (JWST) has enabled the detection of GC populations out to $z \approx 1$. Data from JWST galaxy cluster lensing surveys such as PEARLS \citep{PEARLS}, UNCOVER \citep{UNCOVER}, and CANUCS \citep{CANUCS} revealed point sources around distant BCGs. Early work investigating these sources in SMACS J0723.3-7327 confirmed that they are likely GCs and ultra-compact dwarf galaxies (UCDs) \citep{Faisst2022}. Samples of GC systems now include RXJ~2129 at $z = 0.234$ \citep{Keatley2025}, Abell~2744 at $z=0.308$ \citep{2024Harris}, PLCK~G165.7+67.0 at $z=0.35$ \citep{Hinrichs2026}, 
MACS~J0416.1-2403 at $z=0.397$ \citep{Berkheimer2026}, MACS~J0417.5-1154 at $z=0.443$ \citep{2025Harris}, and El Gordo at $z=0.87$ \citep{2025HarrisReinaCampos}.

Prior to JWST, models of GC evolution could only be evaluated by comparison to observations of GCs in the local universe. With new observations of GC systems at lookback times spanning over 7 billion years of cosmic history, we can test the models of GC disruption at various stages in their evolution. This presents a much more rigorous test of GC evolution models, and thus an excellent opportunity to refine our understanding of GC formation and disruption.

In this work, we present predictions for the GC populations of BCGs using a hierarchal model for GC evolution and disruption. Where possible, we discuss how our results relate to recent JWST observations of GC systems in distant galaxy clusters. In \S\ref{sec:catalogs}, we describe the model for GC formation and evolution and the resulting GC catalogs used in this investigation. In \S\ref{sec:evolution}, we present quantitative predictions for the evolution of the GC mass function and surface density profile, and discuss the significance of massive satellite systems. In \S\ref{sec:metallicity} we describe the metallicity distribution and mass-metallicity relation for model GCs of giant galaxies. In \S\ref{sec:in-situ_ex-situ}, we investigate the spatial distribution of in-situ and ex-situ formed GCs around massive host galaxies and test methods for classifying GCs based on their observable properties. We discuss the interpretation of our results and compare them with observations in \S\ref{sec:discussion}. We summarize our finding in \S\ref{sec:conclusion}. 

\section{Globular cluster model catalogs}
\label{sec:catalogs}

We use the globular cluster formation and evolution model presented in \cite{ChenGnedin2024}. This model post-processes snapshots of cosmological simulations to produce a catalog of GCs.

\subsection{Background Cosmological Simulations}

We used snapshots of the hydrodynamic simulations Illustris TNG100-1 and TNG300-1 (\citealt{nelson2021}, \citealt{Pillepich2017}, \citealt{Springel2017}, \citealt{Nelson2017}, \citealt{Naiman2018}, \citealt{Marinacci2018}, hereafter TNG100 and TNG300).

\begin{table*}[]
    \centering
    \caption{Observational limits on GC populations of distant galaxy clusters. The columns are: cluster name; redshift; lookback time; NIRCam band; limiting absolute magnitude of GC for 50\% completeness including K-corrections; corresponding GC mass limit, see \S\ref{sec:mass_func_discussion}; limit of the observed radial extent from the galaxy cluster center; the detection reference. For MACS~J0416.1-2403, Abell~2744, and El~Gordo the limiting radial extent was judged from plots of the spatial distribution of GC detections, otherwise the reported radial profile was used. PLCK~G165.7+67.0 and El~Gordo are merging clusters, we list the radial extent in both of their subclusters labeled by their relative cardinal directions.} 
    \label{tab: obs_limits}
    \begin{tabular}{lccccccl}
    \toprule
        Cluster name & $z$ & $T_{L}$ & Band & $M_{\rm AB,lim}$ & $\log M_{\rm lim}$ & $R_{\rm lim}$ & Reference \\
        & & (Gyr) & & (mag) & ($\Msun$) & (kpc) \\
        \hline
        RXJ 2129 & 0.234 & 2.9 & F150W & --10.7 & 6.0  & 330 & \citealt{Keatley2025} \\
        Abell 2744 & 0.308 & 3.5 & F150W & --11.1 & 6.1  & 800 & \citealt{2024Harris} \\
        PLCK G165.7+67.0 & 0.350 & 3.9  & F200W & --11.2 & 6.2  & 150 (SW) 150 (NE)& \citealt{Hinrichs2026} \\
        MACS J0416.1-2403 & 0.397 & 4.4 & F200W & --10.6 & 5.9 & 300 & \citealt{Berkheimer2026} \\
        MACS J0417.5-1154 & 0.443 & 4.7 & F150W & --12.1 & 6.5 & 300 & \citealt{2025HarrisReinaCampos} \\
        El Gordo & 0.870 & 7.4 & F150W & --13.9 & 7.2 & 100 (SE) 400 (NW)
        &  \citealt{2025Harris} \\
    \bottomrule
    \end{tabular}
\end{table*}

The TNG simulations are run with the moving mesh hydrodynamic code \textsc{AREPO} \citep{Springel2010} and adopt a cosmology consistent with \cite{PlanckCollaboration2016} results: $\Omega_{\rm b} = 0.0486$, $\Omega_{\rm m} = 0.3089$, $\Omega_{\rm \Lambda} = 0.6911$, $h = 0.6774$, $\sigma_{8} = 0.8159$, and $n_s =  0.9667$ \citep{nelson2021}. TNG100 is initiated with $1820^3$ dark matter particles and gas cell in a $106.5^3$ Mpc comoving box. The average dark matter particle mass is $7.5 \times 10^{6}\Msun$ and the average gas cell mass is $1.4\times 10^6 \Msun$ \citep{nelson2021}. TNG300 is initiated with $2500^3$ dark matter particles and gas cells in a $302.6^3$ Mpc comoving box. The average dark matter particle mass is $5.59 \times 10^{7}\Msun$ and the average gas cell mass is $1.1\times 10^7 \Msun$ \citep{nelson2021}. Halos in the TNG simulations are found using the Friends-of-Friends method, and subhalos are identified with the algorithm \textsc{subfind} \citep{Springel2001}. Hereafter we use `galaxy' to refer to TNG `subhalos'.

We modeled the GC populations of two BCG-like galaxies in TNG100 and one BCG-like galaxy in TNG300. TNG100 has higher mass and spatial resolution, which allows us to better estimate the tidal field along cluster trajectories. TNG300 produces more massive halos, but with coarser resolution. The halo mass ($M_h \equiv M_{200c}$) of two galaxy clusters from TNG100 is approximately the same, $\log(M_h/\Msun) = 14.576$ (subfindID at $z = 0$: 0) and $\log(M_h/\Msun) = 14.581$ (subfindID at $z = 0$: 17185), which is small compared to the JWST clusters where GC populations have been observed, typically $\log(M_h/\Msun) > 15$. The galaxy cluster from TNG300 with $\log (M_h/\Msun) = 15.186$ (subfindID at $z = 0$: 0) is closer in mass to the observed clusters. Hereafter we refer to each model realization as Simulation-subfindID: TNG100-0, TNG100-17185, and TNG300-0.

Our model is limited by the mass resolution of TNG100 and TNG300. Reversing the model prescription for GC formation, we can estimate the minimum halo mass $M_h$ required to form GCs at a reference epoch $z = 4.0$ near the expected peak of GC formation rate \citep[e.g.,][]{Choksi2019}. To form GCs with $M = 10^{5} \Msun$ requires $M_h \sim 10^{9.6} \Msun$. TNG100 easily resolves halos of this size with about 500 dark matter particles. TNG300 also resolves such halos, although only with 60 particles. We allow the model to sample the initial cluster mass function (ICMF) down to $M = 10^4 \Msun$. To form GCs of this mass at $z = 4.0$ requires $M_h \gtrsim 10^{9.0}\Msun$. In TNG100 halos at the lower end of this threshold are resolved with about 100 particles, but in TNG300 they contain only about 10 dark matter particles. Thus, at the low mass-end we anticipate some noise in the GC formation process in TNG300 because of its limited resolution. Partly to mitigate this effect, we restrict our quantitative analysis to GCs above a limiting mass $\log(M_{\rm lim}/\Msun) = 5.0$ (see \S\ref{sec:obs}). It is still a much lower limit than is presently observable in the high-redshift systems.

\subsection{Correspondence to Observations}
\label{sec:obs}

For each model realization, we track the evolution of the GC population across 6 representative epochs: $z$ = 1.0, 0.5, 0.4, 0.3, 0.2, and 0.0. These epochs are chosen to correspond roughly with the observed GC populations: redshift $z\approx 0.2$ corresponds to RXJ~2129 at $z = 0.234$, redshift $z\approx 0.3$ corresponds to Abell~2744 at $z=0.308$, redshift $z\approx 0.4$ corresponds to MACS~J0416.1-2403 at $z=0.397$, and redshifts $z=0.5$ and $z=1.0$ are somewhat earlier than MACS~0417.5-1154 and El~Gordo, respectively. At each epoch we record the mass and the projected distance from the host galaxy for each GC. The metallicity of each GC is fixed at the time of its formation. The evolution of the metallicity distribution can be recovered by removing GCs disrupted prior to each epoch.

Observational samples are limited by the depth of their point-source detection and background subtraction. In distant galaxy clusters, only the high-mass end of the GC mass function is observable, and the radial extent of the observed populations is limited by the NIRCam field of view and varies between studies. Table~\ref{tab: obs_limits} gives the limiting mass and radius of recent observational studies of distant GC populations. Our model only produces GCs up to a few times $10^{7}\Msun$ (see \S\ref{sec:model}), while the masses of observed GCs in distant galaxy clusters are close to the upper limit of our model. The model explicitly does not include UCDs. Throughout this work, we apply a limiting mass threshold $M_{\rm lim}$ and a limiting radial extent $R_{\rm lim}$ to our model GC populations for consistency with observations. For the initial analysis, we chose $\log (M_{\rm lim}/\Msun) = 5.0$ and $R_{\rm lim} =1$~Mpc, which are weaker constraints than typical observational limitations. This allows for better statistics for investigating our model's predictions. Later in \S\ref{sec:metallicity} we consider much stricter limits on GC mass.

\begin{figure*}
    \centering
    \includegraphics[width=\linewidth]{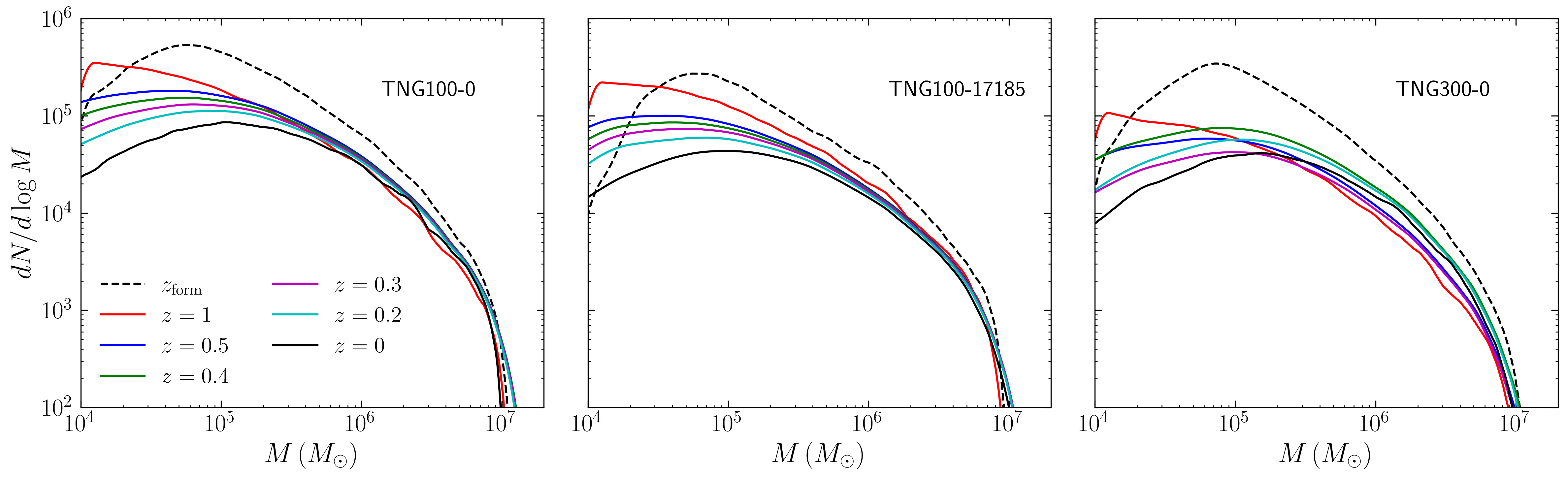}
    \caption{GC mass functions for all three model realizations at different epochs ($R_{\rm lim} = 1$~Mpc). The TNG300 distribution is scaled down by the ratio of the galaxy cluster mass to the mass of the TNG100 galaxy clusters. The distribution of all initial masses (labeled $z_{\rm form}$) is shown as dashed line.}
    \label{fig: mass function}
\end{figure*}

\subsection{Model of GC Formation and Evolution}
\label{sec:model}

The model in \cite{ChenGnedin2024} used for this work is a modification of the previous versions developed in a series of work by \cite{Muratov2010}, \cite{Li2014}, \cite{Choksi2018}, and \cite{ChenGnedin2022}. The current model predicts the spatial distribution by tagging simulation particles as tracers of GC orbits in the host galaxy. Below we provide a brief description of the model. 

We model GC formation and evolution in three steps: 1) cluster formation, 2) cluster sampling and particle assignment, and 3) cluster evolution. GC formation occurs during rapid growth of the host galaxy -- it is triggered when the specific mass accretion rate of a selected halo $\dot{M_{\rm h}}/M_{\rm h}$ exceeds the model threshold parameter $p_3$. We apply the empirical stellar mass-halo mass relation (SMHM) relation \citep{Behroozi2013} to compute the galaxy stellar mass at a given redshift, and the empirical gas mass-stellar mass relation \citep{ChenGnedin2024} to compute the gas mass from the stellar mass. The total mass of a GC population forming at a given epoch is given by the linear gas mass-cluster mass relation $M_{\rm tot} = 1.8 \times 10^{-4}\, p_2\, M_{\rm gas}$ \citep{Kravtsov2005}, where $p_2$ is another model parameter. To determine the GC metallicity, we first find the host galaxy metallicity using the empirical mass-metallicity relation (MZR) with its scatter modeled as a Gaussian process, and then assign the GCs the metallicity of the host galaxy at its time of formation with an additional scatter of 0.2~dex \citep{ChenGnedin2024}. To obtain the masses of the individual clusters we sample from the \cite{Schechter1976} ICMF with an exponential cutoff at $M_c = 10^{7}\Msun$ and a minimum mass $M_{\rm min} = 10^4 \Msun$. Only clusters with $M > 10^4 \Msun$ at $z= 1$ are included in the final catalogs because lower mass GCs are likely to be disrupted in less than 1 Gyr. The individual clusters are then assigned to young stellar particles within 3~kpc of the galaxy center in the relevant simulation snapshot. Finally, we evaluate the GC mass loss due to stellar evolution and tidal disruption based on the local tidal field along the GC's orbit. The mass loss due to stellar evolution is modeled as an instantaneous reduction of 45\% at formation following \cite{Gieles2023}. This is a valid approach given that the timescales for stellar evolution are much shorter than GC lifetimes. The tidal disruption is modeled using a multivariate power-law function of the cluster initial mass, current mass, and the strength of tidal field \citep{Gieles2023}. The limited spatial and temporal resolution of the simulation may lead to an underestimate of the tidal field strength. To correct for this underestimation, we use a third adjustable model parameter $\kappa$. The current model parameters ($p_2$, $p_3$, $\kappa$) = (18, 0.5 $\mathrm{Gyr^{-1}}$, 1.5) were calibrated with observations of the Milky Way and M31 in \cite{ChenGnedin2024}. The model has been validated for Milky Way-sized halos, the results here are assuming the model generalizes to giant galaxies.

\section{Evolution of Globular Cluster Populations from redshift 1}
\label{sec:evolution}

We begin by investigating the evolution of the GC mass function and surface density profile in each model realization. We restrict the radial extent for the mass function with $R_{\rm lim} = 1$~Mpc, and we set a minimum mass $\log (M_{\rm lim}/\Msun) = 5.0$ for the surface density profile. For the mass function, we did not apply a limiting mass since we are interested in the full mass distribution. Likewise, we did not apply a radial limit for the surface density profile. 

To visualize the distribution functions, we use Kernel Density Estimators (KDEs) with an Epanechnikov kernel. The KDE function with bandwidth $h$ is defined as
\begin{equation}
    f(x) =  \sum_i^N {1\over h} K_E{\left(\frac{x-x_i}{h}\right)},
    \label{eq: KDE}
\end{equation}
where $K_E(u)$ is the Epanechnikov kernel function \(K_{E}\, (u) = \frac{3}{4} (1-u^2)\) for \(|u| \le 1\), and \(K_{E}\, (u) = 0\) otherwise. For visualization purposes $h$ is selected using Sliverman's rule of thumb:
\begin{equation}
     h_0 = 0.9 \min\left(\sigma, \frac{IQR}{1.34}\right) \; N^{-1/5},
     \label{eq: h_0}
\end{equation}
where IQR is the interquartile range of the distribution, and $N$ is the number of objects. This selection criterion was designed for a Gaussian kernel but works well for finding a balanced bandwidth for an Epanechnikov kernel. We set a minimum value $h_{\rm min} = 0.1$ to prevent under-smoothing.

\begin{figure*}
    \centering
    \includegraphics[width=\linewidth]{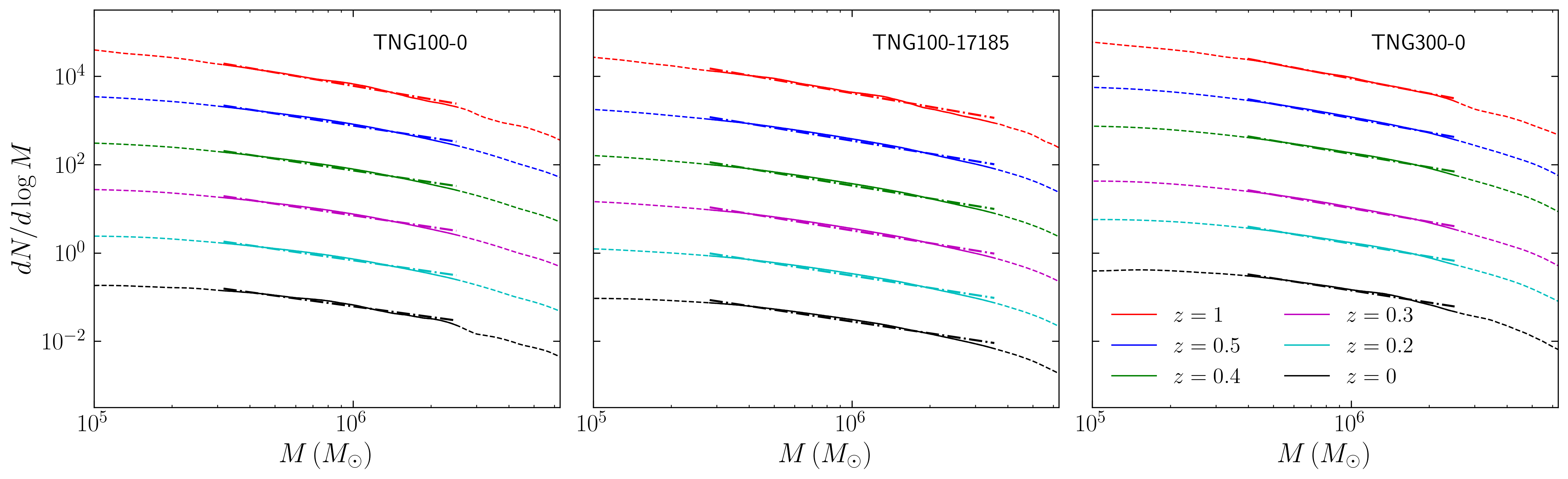}
    \caption{A comparison of the mass functions to the best-fit power laws using MLE (dash-dot lines), for $R_{\rm lim} = 1$~Mpc. Each epoch is offset in normalization for clarity. The solid part of the curve (fixed across all epochs) represents the region where the power law was fit.}
    \label{fig: mass function + fit fixed range}
\end{figure*}

\begin{figure}
    \centering
    \includegraphics[width=\linewidth]{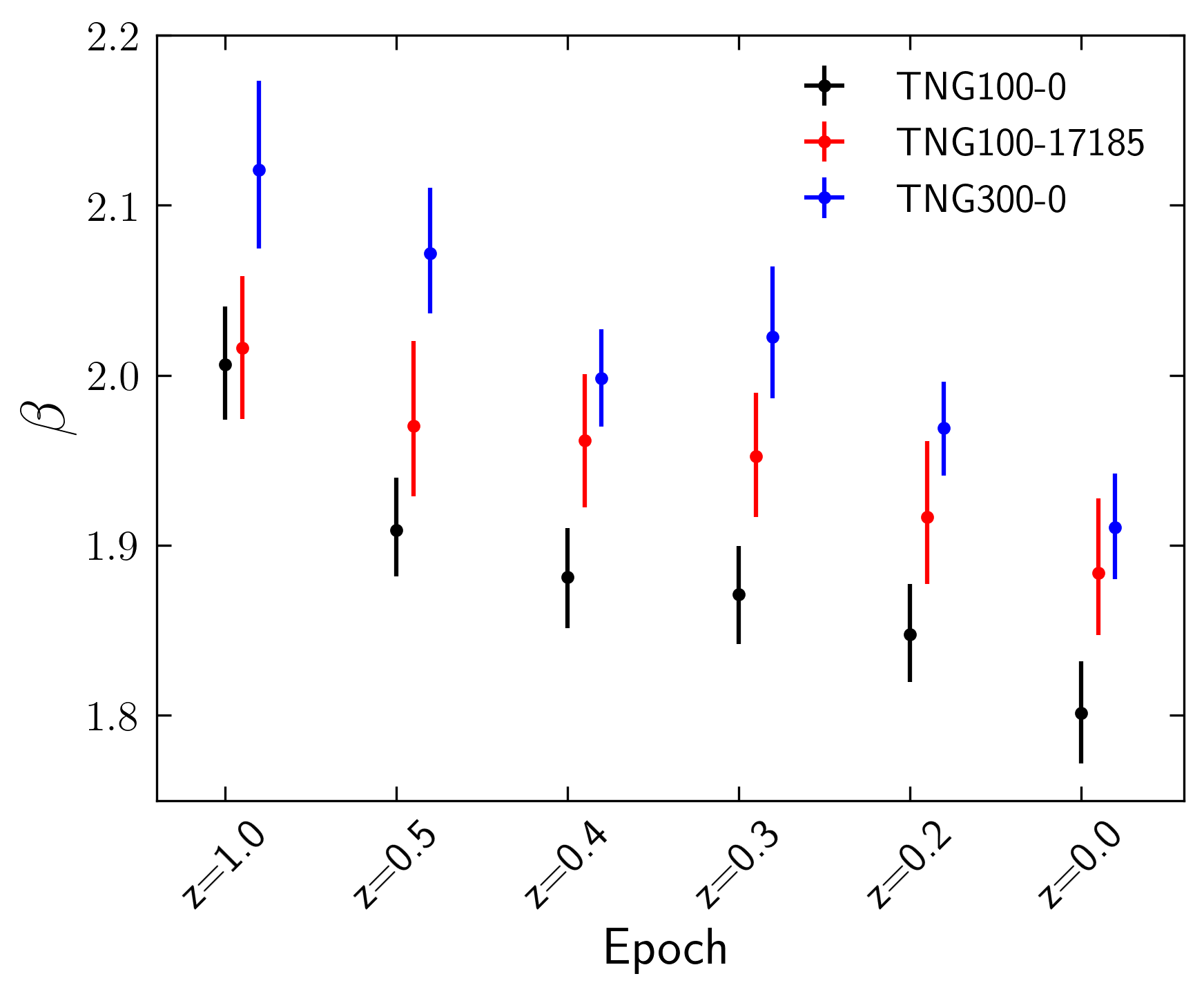}
    \caption{Power law slopes of the GC mass function, obtained using MLE. Error bars represent the 16th and 84th percentiles. The slopes become shallower over time as low-mass GCs are preferentially disrupted.}
    \label{fig: mass_function_slopes}
\end{figure}

\subsection{Mass Function}

The globular cluster mass function (MF) measures the number of GCs per unit logarithmic mass interval. Figure~\ref{fig: mass function} displays the MF of each model realization over 6 epochs of evolution. The dashed curve gives the initial mass function at the redshift of formation. The model catalogs only include clusters that had mass $M>10^4\Msun$ at $z=1$, which explains the downturn of the plotted (but not actual) initial MF due to some low-mass GC already being disrupted by $z=1$.

In the TNG300-0 sample, the MF grows until $z=0.4$ for GCs less affected by dynamical disruption, with $M > 10^5\Msun$. This growth is not due to the formation of new GCs. Instead, it is due to the accretion of giant satellite systems, which we discuss in more detail in \S\ref{sec:satellites}.

In nearby galaxies the MF of young star clusters is well described by a power law \citep[e.g.,][]{Chandar2026}. As a result of dynamical disruption, the low-mass end of the GC MF is depleted, but the high-mass end approximately retains the original shape. Therefore, we attempt to fit the high-mass end of the model cluster MF by a power law:
\begin{equation}
    \dfrac{dN}{dM} = C M^{-\beta}
    \label{eq: mass func power law}
\end{equation}
where $C$ is the normalization and the expected slope $\beta \approx 2$. This functional form applies only above a certain minimum mass, which we determine empirically.

We explored two different approaches for fitting the MFs. The primary method used in this work is maximum likelihood estimation (MLE). For a distribution that follows $dN/dx = Cx^{-\beta}$ between $x_{\rm min}$ to $x_{\rm max}$, the log-likelihood function is
\begin{equation}
    \ln \mathcal{L} =  N\ln{b} - N\ln(x_{\rm max}^b - x_{\rm min}^b)+ (b-1)\sum_{i=1}^N \ln x_i.
    \label{eq: log likelihood MLE}
\end{equation}
Here, $b = 1-\beta$ and $N$ is the size of the sample. To find the value of $b$ that maximizes the likelihood function, we solved $\partial \ln \mathcal{L}/\partial b = 0$ using the \textsc{Scipy} implementation of Brent's method for root finding, \texttt{brentq} \citep{Brent1973}. From the best-fit value of $b$ we can calculate $\beta = 1-b$ and $C = N b/(x_{\rm max}^b - x_{\rm min}^b)$.  

We evaluated the quality of the fit using a one-sample Kolmogorov-Smirnov test, whose $p$-values greater than 0.01 were considered sufficient. For a robust estimate of uncertainties given the large size of the GC catalogs, we used block bootstrap, in which we divided the distribution into blocks of $N=1000$ clusters and randomly sampled a set of blocks with replacement to create a bootstrap sample. For each bootstrap sample, we solve for $b$ and $C$. After 1000 bootstrap iterations, we calculate the 16th and 84th percentiles of $b$ to estimate its uncertainty with a $68\%$ confidence level. 

We also considered an alternative binning-based method, which involves approximating $dN/d\log M$ with a KDE and performing linear regression between $\log_{10}({dN/d\log M})$ and $\log M$.  In terms of $\log M$, the power law becomes
\begin{equation}
    \frac{dN}{d\log M} =  \dfrac{dN}{dM}M\ln(10) = C\ln(10) M^{-\beta +1}.
    \label{eq: mass func log}
\end{equation}
Taking the base 10 logarithm,
\begin{equation}
    \log \left(\dfrac{dN}{d\log M}\right)= \log(C\ln(10)) + (1-\beta)\log M.
\end{equation}
Thus, the slope of this relationship is $1-\beta$. We defined the following function for the log-likelihood
\begin{equation}
    \ln \mathcal{L} =  \sum_i^N -(y_i - a -bx_i)^2,
    \label{eq: binning log likelihood}
\end{equation}
where $a$ and $b$ are the intercept and slope parameters. We minimized $- \ln \mathcal{L}$ with \textsc{Scipy} \texttt{minimize} to obtain the best-fit values for $a$ and $b$. Similarly to MLE, we estimate uncertainties using block bootstrap with 1000 iterations.

To validate our fitting methods, we tested them on a mock power-law distribution. We generate the mock distribution using the transformation method which uses the cumulative distribution function (CDF) and its inverse to convert uniformly distributed random numbers into the target distribution. The CDF for a power law between $x_{\rm min}$ and $x_{\rm max}$ is
\begin{equation}
    F(x) = \dfrac{x^{1-\beta}- x_{\rm min}^{1-\beta}}{x_{\rm max}^{1-\beta}- x_{\rm min}^{1-\beta}}.
    \label{eq: CDF mock data}
\end{equation}
We used \textsc{Scipy}'s \texttt{UnivariateSpline} routine without smoothing to create an inverse interpolating function which maps the values of $F^{-1}(x)$ to $x$. We input random numbers between 0 and 1 drawn from a uniform distribution into our interpolating function. This results in a distribution with the desired power law $\beta$. We used a sample size $N = 100,000$ which is comparable to the size of the model catalogs ($N = 132,118$ for TNG100-0, $N = 59,478$ for TNG100-17185, and $N = 345,755$ for TGN300-0). We created a mock distribution with an analytical power law $\beta = 2$ in the range $10^{4}< x < 10^{7}$. These parameters were selected to be similar to the GC mass function. The binning routine recovered a slope $\beta = 1.995^{+0.009}_{-0.008}$ and MLE recovered a slope $\beta = 2.001^{+0.003}_{-0.003}$. Thus, both routines successfully matched the input slope within uncertainties. We favor the use of MLE because it recovers $\beta$ with smaller uncertainties and does not rely on binning the data.

\begin{table}[htbp]
\centering
\small
\caption{Power law fitting ranges ($\min-\max$) for the mass function (MF) and surface density profile ($\Sigma$).}
\label{tab:fit-ranges}
\begin{tabular}{lcccc}
\toprule
& & \multicolumn{2}{c}{MF} & $\Sigma$ \\
\cmidrule(lr){3-4} \cmidrule(lr){5-5}
Halo ID & $z$ & Variable & Fixed & Fixed \\
& & \multicolumn{2}{c}{$\log{(M/\Msun)}$} & $\log{R}$~(kpc) \\
\midrule
\multirow{6}{*}{TNG100-0} 
  & 1.0 & 5.40--6.45 & \multirow{6}{*}{5.50--6.40} & \multirow{6}{*}{1.1--1.8} \\
  & 0.5 & 5.50--6.40 & & \\
  & 0.4 & 5.55--6.50 & & \\
  & 0.3 & 5.45--6.40 & & \\
  & 0.2 & 5.45--6.40 & & \\
  & 0.0 & 5.50--6.40 & & \\
\midrule
\multirow{6}{*}{TNG100-17185} 
  & 1.0 & 5.45--6.55 & \multirow{6}{*}{5.45--6.55} & \multirow{6}{*}{1.0--2.0} \\
  & 0.5 & 5.45--6.60 & & \\
  & 0.4 & 5.45--6.55 & & \\
  & 0.3 & 5.45--6.60 & & \\
  & 0.2 & 5.40--6.60 & & \\
  & 0.0 & 5.40--6.55 & & \\
\midrule
\multirow{6}{*}{TNG300-0} 
  & 1.0 & 5.55--6.60 & \multirow{6}{*}{5.60--6.40} & \multirow{6}{*}{1.1--1.9} \\
  & 0.5 & 5.50--6.40 & & \\
  & 0.4 & 5.55--6.40 & & \\
  & 0.3 & 5.50--6.40 & & \\
  & 0.2 & 5.55--6.40 & & \\
  & 0.0 & 5.60--6.40 & & \\
\bottomrule
\end{tabular}
\end{table}

The inferred power law parameters can be sensitive to the range used for the fit. We tested two methods to select the region of the fit: a variable range across redshift and a fixed range at all redshifts. In both cases, we searched between $5.4 <\log{(M/\Msun)} < 6.6$ and found the largest ranges that gave the Kolmogorov-Smirnov test $p$-value $> 0.01$ between the MLE fit and the actual MF. A different fixed range was used for each BCG halo. The selected fitting ranges are listed in Table~\ref{tab:fit-ranges}. We find that the difference between using a variable range vs. a fixed range is minor. The mean absolute error between the variable range and fixed range slopes is 0.022, the mean absolute error relative to uncertainties is 0.237, and the mean difference is $-0.005$. All mean values are averaged over epoch and model realizations. Given the limited effect of the variable range, we focus on our results for a fixed range to simplify the interpretation of the evolution over time. Note that the MF fitting ranges extend below typical observational limits of the high-redshift BCGs. The analysis in this section focuses on a theoretical understanding of the MF evolution in the giant GC systems. 

Figure~\ref{fig: mass function + fit fixed range} compares the MLE power law fit over a fixed range with the actual mass functions. An offset is added on the $y$-axis to prevent overlap between lines and improve readability. The fits are very accurate for the chosen mass range.

Figure~\ref{fig: mass_function_slopes} presents the evolution of the slopes measured with MLE and a fixed mass range. All the values are $\approx 2$, consistent with the mass function of young star clusters and with the model ICMF in a given star-forming episode. However, we detect a significant trend in each model realization: the slope becomes shallower over time. This can be explained by the preferential disruption of low-mass clusters, which is expected to produce a progressively shallower MF. There was no significant difference between the use of the MLE or binning methods. The mean absolute error between the MLE power law slopes and the binning power law slopes is 0.024, the mean absolute error relative to uncertainties is 0.264, and the mean difference is $-0.024$.

\begin{figure*}
    \centering
    \includegraphics[width=\linewidth]{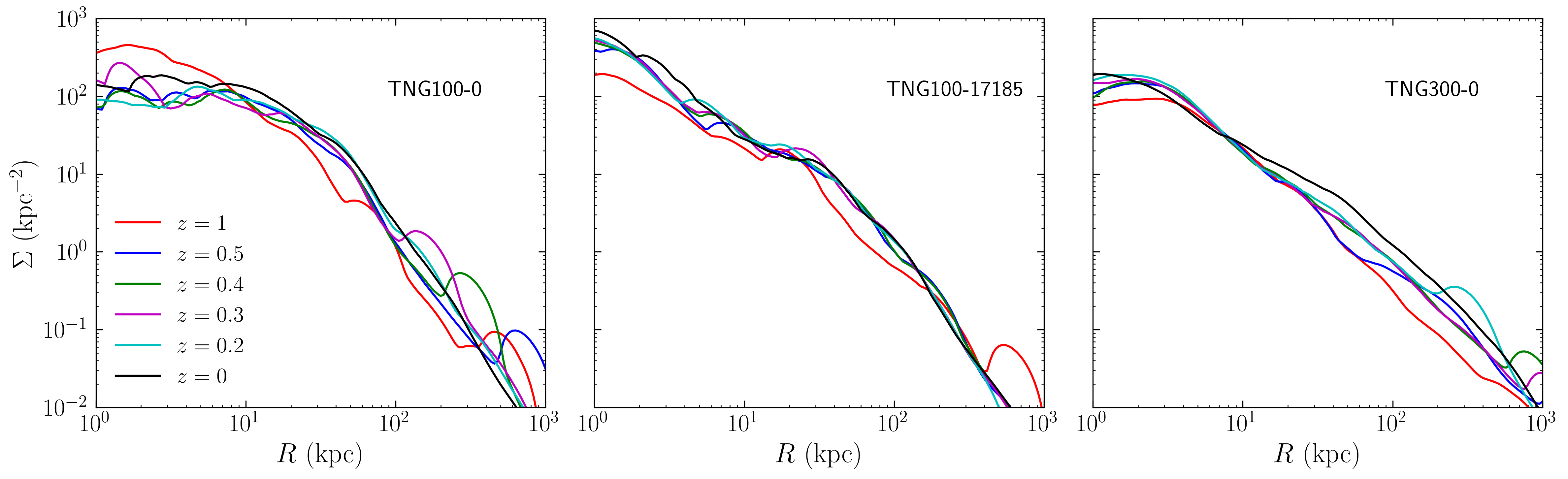}
    \caption{The surface density profiles for each model realization, scaled by relative galaxy cluster mass as in Figure~\ref{fig: mass function}, for $\log(M_{\rm lim}/\Msun)=5.0$.}
    \label{fig:surf den}
\end{figure*}

\subsection{Surface Density Profile}

\begin{figure*}
    \centering
    \includegraphics[width=\linewidth]{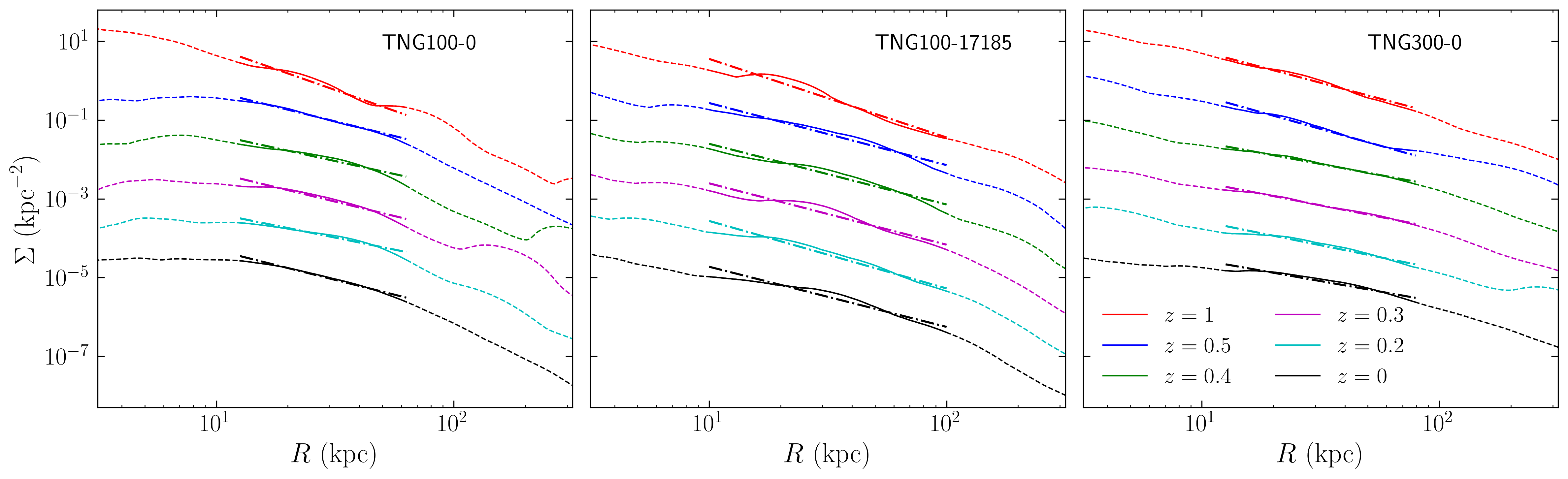}
    \caption{A comparison of the surface density profiles to the best fit power laws (dash-dot lines), for $\log(M_{\rm lim}/\Msun) =5.0$. Each epoch is offset in normalization for clarity. The solid part of the curve (fixed across all epochs) represents the region where the power law was fit.}
    \label{fig: surf den + fit}
\end{figure*}

\begin{figure}
    \centering
    \includegraphics[width=\linewidth]{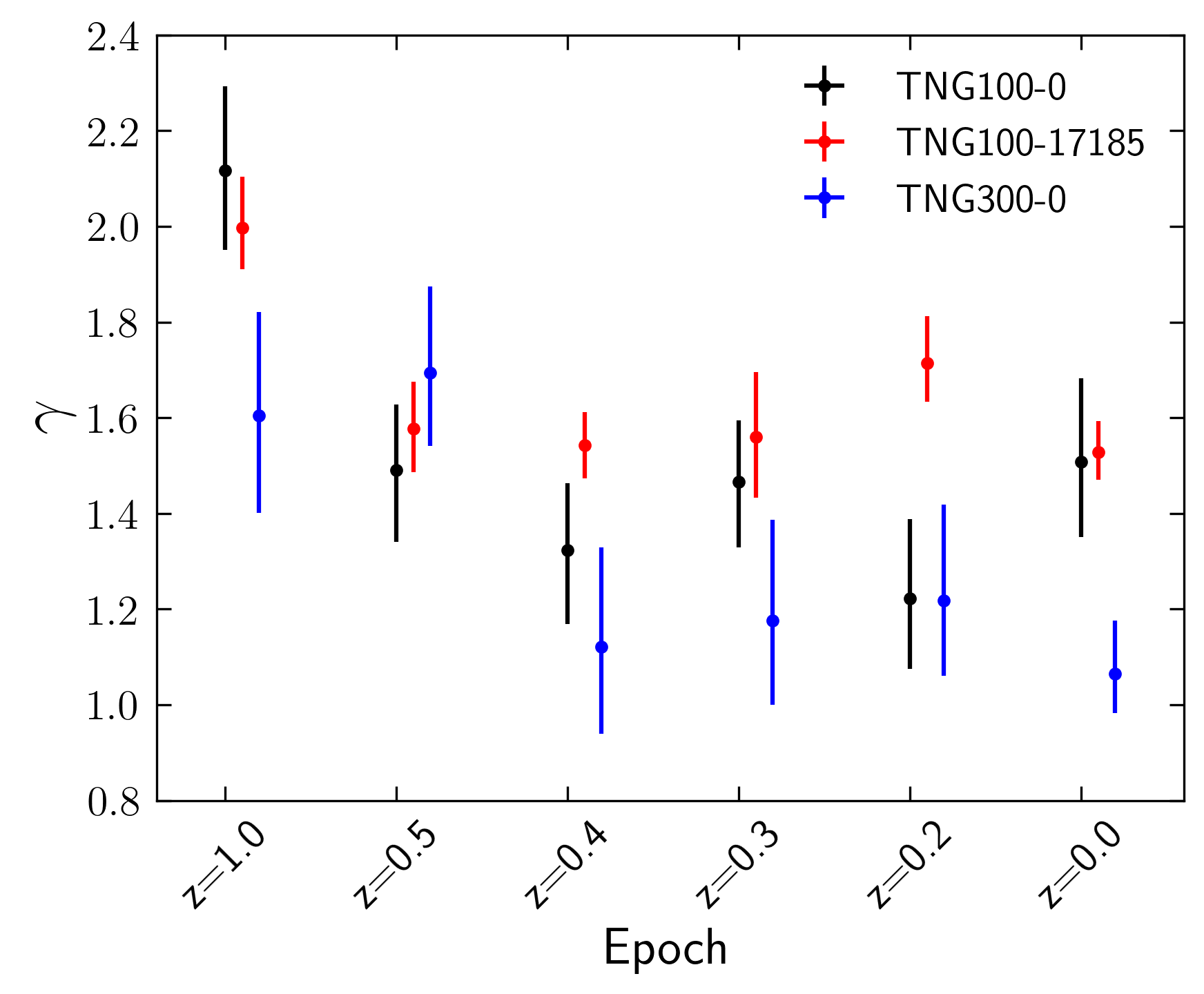}
    \caption{Slopes of the surface density power-law fit with MLE. Error bars represent the 16th and 84th percentiles. The power law is steeper at earlier times corresponding to a tighter clustering of GCs around their host galaxy.}
    \label{fig:SDP_slopes}
\end{figure}

The surface density ($\Sigma$) profile measures the 2D spatial distribution of GCs around their host galaxies. It is defined as
\begin{equation}
    \Sigma = \dfrac{1}{2\pi R} \dfrac{dN}{dR} =
    \dfrac{1}{2\pi R^2 \ln(10)} \dfrac{dN}{d\log R}.
    \label{eq: surf den}
\end{equation}
The projected radius $R$ is calculated in the model as the average of three projections in the simulation coordinates $R = ((x^2+y^2)^{1/2} + (y^2+z^2)^{1/2} + (x^2+z^2)^{1/2})/3$.

Figure~\ref{fig:surf den} presents the surface density profiles of the GC populations of each model realization over the 6 epochs. To visualize the surface density profile, we compute $dN/d\log R$ using KDE. Visible ``bumps" at $R>100$~kpc are the GC systems of massive satellite galaxies, distinct from the central BCG system.

Like the mass function, the surface density profile can be described by a power law of the form 
\begin{equation}
    \Sigma = C R^{-\gamma}.
    \label{eq: surf den power law}
\end{equation}
This is the same functional form as the MF power law, so we applied the same fitting routines to determine the slope $\gamma$.  

We aimed to fit the power law in the region between 10-100~kpc to exclude the BCG core and satellite galaxies. Even with these restrictions, the surface density profiles are not smooth in this region. The profiles do not closely follow a single power law, and the Kolmogorov-Smirnov $p$-values of the fit are less than the desired 0.01. We cannot use the condition $p$-value $>0.01$ to select the fitting range. Instead, we selected by eye a fixed range for each BCG which can approximately be described by a power law. Table~\ref{tab:fit-ranges} gives the fitting ranges.

Figure~\ref{fig: surf den + fit} compares the MLE power-law fit to the surface density profiles. Similarly to the mass functions, we applied vertical offsets to improve readability. For TNG100-0 and TNG300-0 the profiles are relatively smooth and the power-law approximation is reasonable. The TNG100-17185 realization shows more structure at progressively larger distances from the center over time.

Figure~\ref{fig:SDP_slopes} shows that the power law slope is higher at $z \ge 0.5$. At later epochs, it appears scattered between $\gamma=1$ and 1.7. We expect the surface density profile to get shallower over time with the accretion of ex-situ GCs at large radii and the tidal disruption of GCs near the galactic center. This is consistent with our results. There is no significant difference between the use of MLE or binning. The mean absolute error between the MLE slopes and the binning slopes is 0.070, the mean error relative to the uncertainties is 0.195, and the mean difference is $-0.049$.

\subsection{The Importance of Satellite Systems} 
\label{sec:satellites}

\begin{figure}
    \centering
    \includegraphics[width=\linewidth]{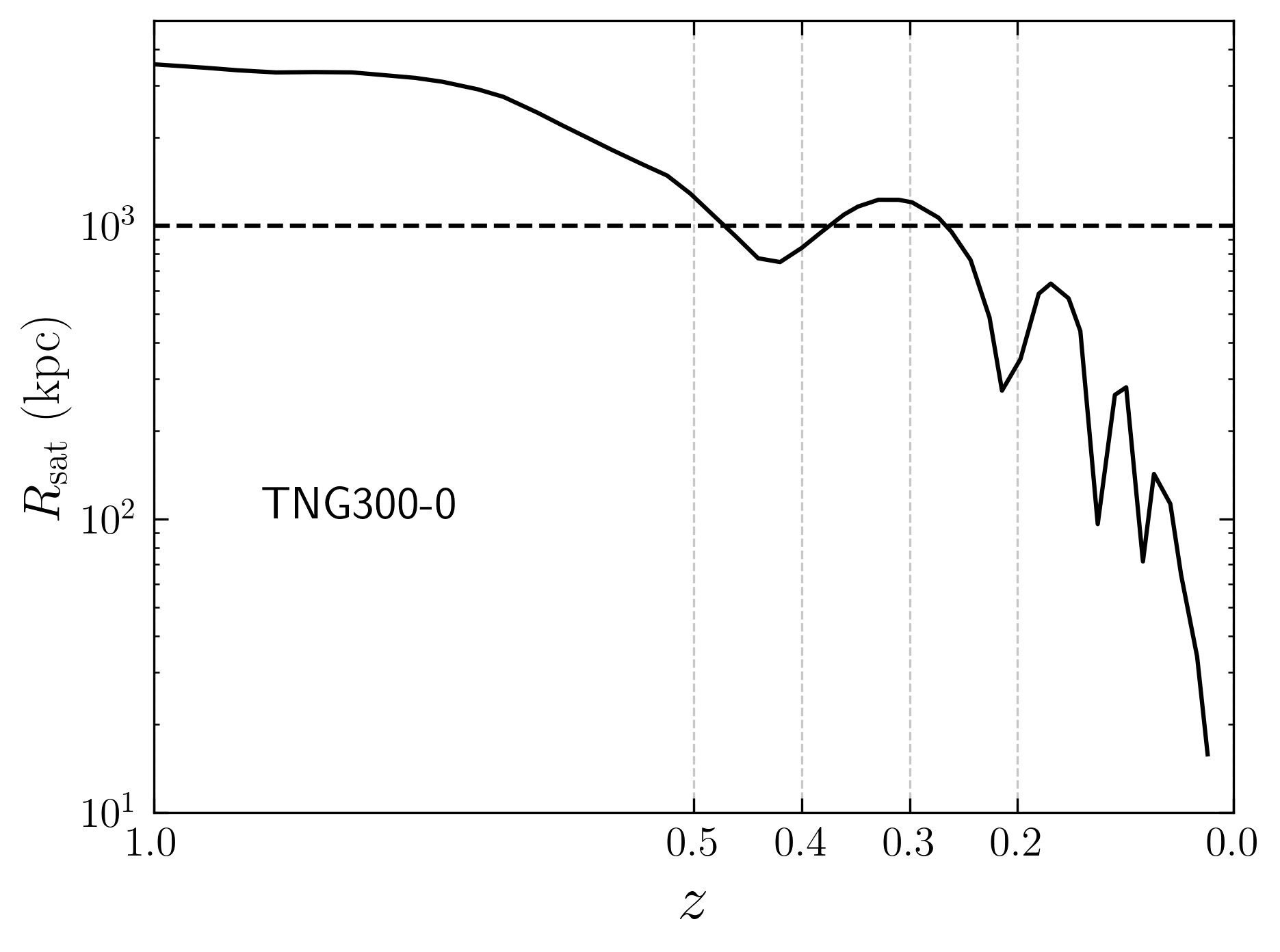}
    \caption{Distance to the most massive satellite of TNG300-0 over time. The dashed horizontal line represents $R_{\rm lim} = 1$~Mpc. The satellite moves in an out of the 1~Mpc region of interest before merging with the BCG.}
    \label{fig:merger_distance_vs_z}
\end{figure}

The contributions of satellite galaxies to the total GC population can be significant. We find this to be particularly true for TNG300-0. Figure~\ref{fig:merger_distance_vs_z} tracks the distance of TNG300-0's most significant satellite from the central halo. The normalization of the mass function for $M>10^5 \Msun$ (where GCs are less affected by disruption) is largest at $z = 0.4$, which corresponds to the epoch when this satellite enters the 1~Mpc region of interest. It contributes $\sim 18,000$ GCs between 400--1000 kpc. At $z= 0.3$, the number of GCs in the same radial range decreases by a similar amount, indicating in agreement with Figure~\ref{fig:merger_distance_vs_z}, that this massive satellite has moved out of the 1~Mpc region. By $z = 0.2$ the satellite galaxy has returned and is located about $200$~kpc from the BCG. It now contributes $\sim13,000$ new GCs. Accordingly, the normalization of the mass function at $z = 0.2$ is the second largest. This satellite system appears as ``bumps" in Figure~\ref{fig:surf den} at $z = 0.4$ and $z = 0.2$. The satellite eventually merges with the BCG at $z \approx 0.1$. 

Both TNG100 realizations of the model are also affected by satellite galaxies, but to a smaller degree. Their satellites contribute at most $5,000$ GCs, and in nearly all epochs, the majority of GCs fall within 100~kpc of the BCG. In TNG100-17185 at $z=1$ a significant satellite galaxy at 400~kpc from the BCG contributes about 1/3 of the total GCs. The difference in the satellite contribution between the TNG100 halos and TNG300-0 suggests that massive satellites are a property of very massive systems with $\log(M/\Msun) \gtrsim 15$. With only one halo realization above this mass, we cannot rule out the possibility that TNG300-0 has uniquely massive satellite systems. 

The effect of satellite galaxies is evident in the half-mass radius of the whole GC system, $R_{\rm e}$. The half-mass radius is the median 2D projected radius of GCs from the host galaxy, which we measured using $\log(M_{\rm lim}/\Msun) = 5.0$ and $R_{\rm lim} = 1$~Mpc. Table~\ref{tab:crossover_r} lists the half-mass radii of the three halo realizations as a function of redshift. We see that $R_{\rm e}$ of TNG300-0 is dramatically larger at $z = 0.4$ and $z = 0.2$, corresponding to the presence of the massive satellite in the 1~Mpc region. The half-mass radius of TNG100-17185 is nearly three times larger at $z = 1$ than at later times due to the massive satellite at 400~kpc. 

If we relax the limiting radius to $R_{\rm lim} = 10^{3.5}$~kpc, the half-mass radius of TNG300-0 exceeds 500~kpc at $z =$0.3--1.0 due to contributions from massive satellite galaxies beyond 1~Mpc. Such satellites are beyond current observable limits and their GCs are too far to belong to the central BCG. We do not use $R_{\rm lim} > 1$~Mpc in our analysis. 

Satellite galaxies within 1~Mpc will likely make an important contribution to the total GC population. These galaxies are often only a few hundred kpc from the BCG, and within the radial scope of observations. Thus, we chose not to exclude their contribution from our investigation.

\begin{figure*}
    \includegraphics[width=\linewidth]{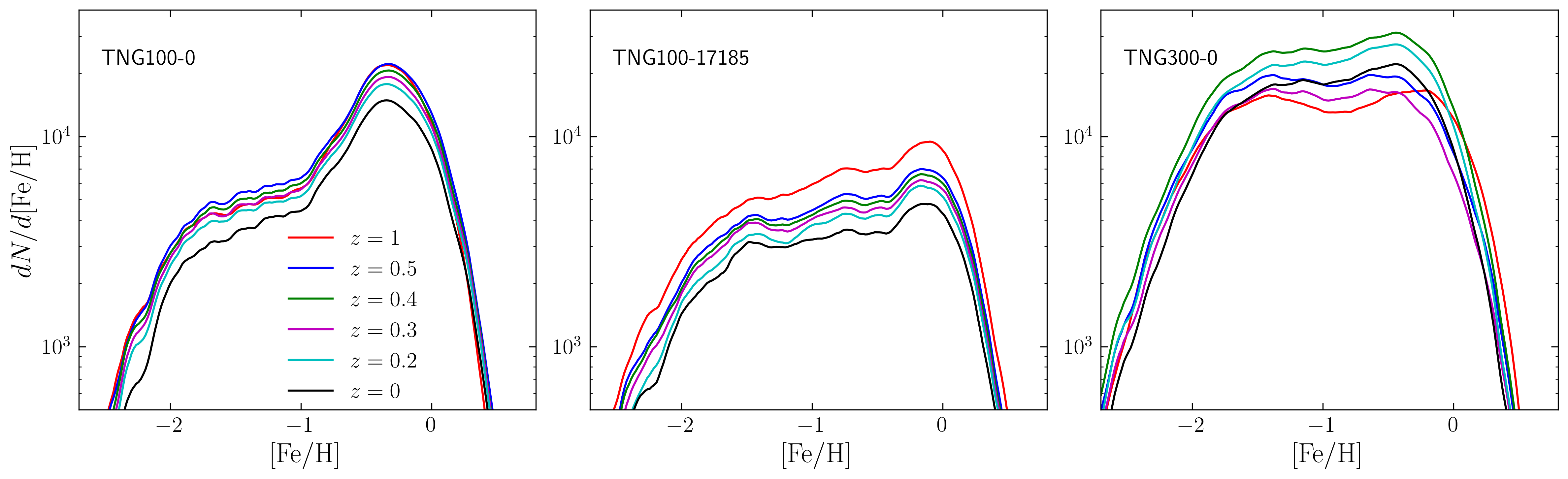}
    \caption{Metallicity distribution of each model realization over time, for $R_{\rm lim} = 1$~Mpc and $\log (M_{\rm lim}/\Msun) = 5.0$.}
    \label{fig:feh_evolution}
\end{figure*}

\section{Metallicity}
\label{sec:metallicity}

Our model assigns the metallicity $\feh$ of GCs at the time of formation. Even though $\feh$ of individual clusters is fixed, the metallicity distribution of the whole GC population may evolve because the sample changes as some GCs are tidally disrupted.
Figure~\ref{fig:feh_evolution} presents the metallicity distribution of each model realization over the 6 epochs. The qualitative shape of the distribution does not change significantly over time; rather, the normalization of the distribution evolves. Each model realization shows a strong high-metallicity peak between $\feh=-0.5$ and 0, and a weaker low-metallicity peak around $\feh=-1.5$. In the TNG100 halos, the normalization of the metallicity distribution decreases over time through dynamical disruption. In TNG300-0, this behavior is modulated by the GC population of its massive satellite galaxy moving in and out of the $1$~Mpc region.

\begin{figure}
    \centering
    \includegraphics[width=\linewidth]{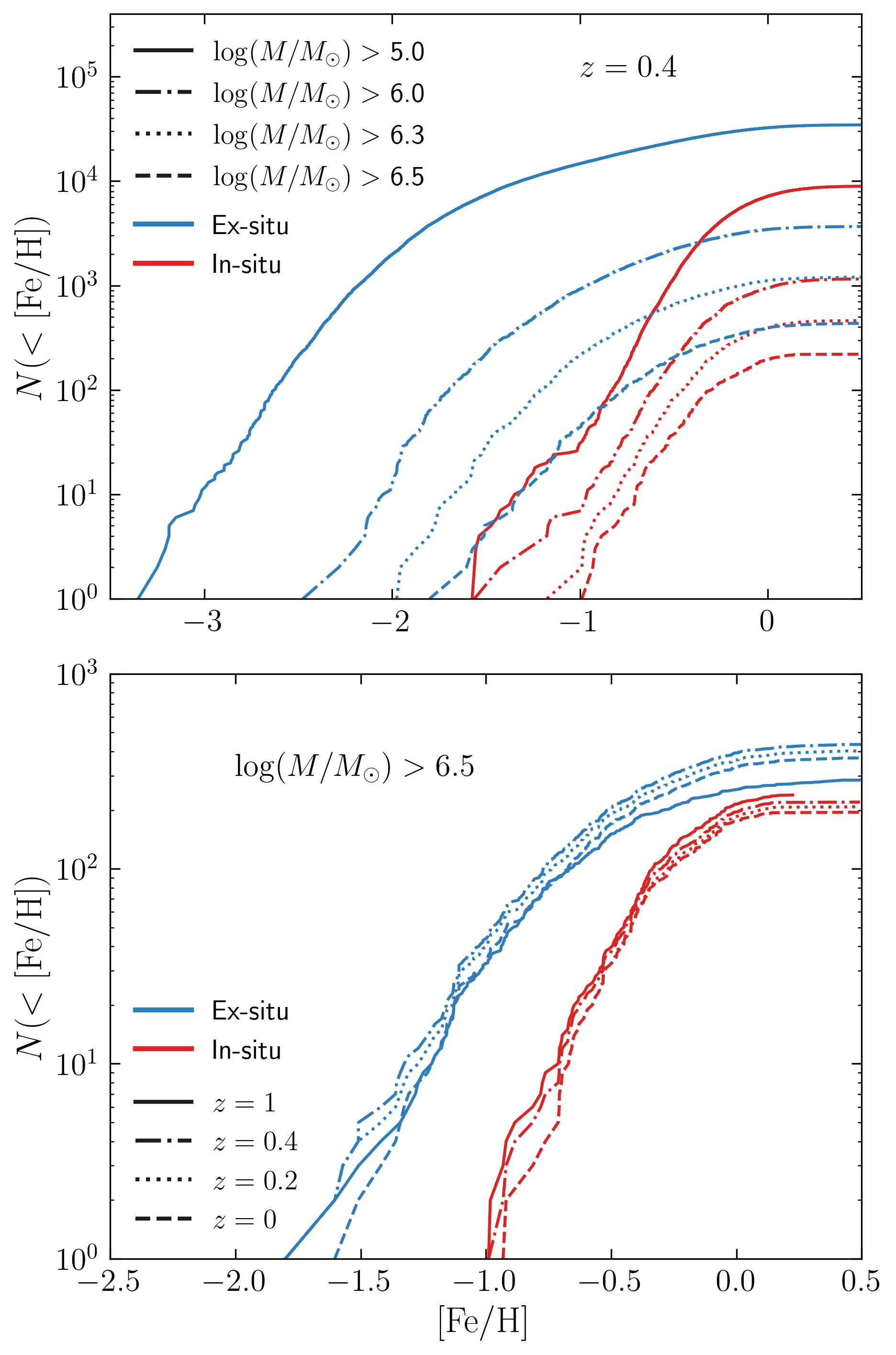}
    \caption{Dependence of the cumulative metallicity distribution on the mass threshold (top) and epoch (bottom). In both panels the radial extent is limited by $R_{\rm lim} = 1$~Mpc. The CDF shifts to higher metallicities with a stricter mass threshold, and the in-situ fraction grows. The shape of metallicity CDF is consistent across epochs.}
    \label{fig:metallicity_cdfs}
\end{figure}

\subsection{Metallicity of In-situ and Ex-situ Clusters} 

Our model distinguishes between GCs formed \textit{in-situ}, in the central host galaxy, and \textit{ex-situ}, in satellite galaxies. Decomposing the GC population into in-situ and ex-situ components provides insight into the formation history of the population. The GC metallicity is determined by the environment in which it formed. Thus, there is a close relationship between the metallicity of a GC and its classification as in-situ or ex-situ. 

Given the similar qualitative shape of the metallicity distribution across model realizations, for the subsequent metallicity analysis, we combined the three halo samples. To avoid the more numerous TNG300-0 dominating the statistics, we scaled down its contribution by the ratio of halo mass (TNG300-0 to TNG100-0). Given the approximately linear scaling between the number of GCs and the mass of the host halo \citep[e.g.,][]{Harris2017a}, this rescaling gives comparable weight to each of the three model realizations.

Figure~\ref{fig:metallicity_cdfs} shows the cumulative distribution functions of in-situ and ex-situ metallicity as a function of mass threshold (top panel) and as a function of time (lower panel). In all cases, the in-situ metallicity CDF starts at higher metallicities than the ex-situ CDF. In-situ clusters, formed in the metal-rich molecular clouds of their host galaxy, are metal-rich relative to the population of ex-situ clusters. 

The lower panel shows the evolution of the in-situ and ex-situ metallicity CDFs with a strict mass threshold $\log(M_{\rm lim}/\Msun) = 6.5$, typical of observed samples. Both CDFs maintain a consistent shape over time. The in-situ population monotonically decreases through disruption. The ex-situ population is modulated by the accretion of satellite galaxies. The ex-situ population is largest at $z=0.4$ and 0.2 because of the contribution of the massive satellite in TNG300-0. The high metallicity ex-situ population is significantly larger at $z = 0$ than $z = 1$ likely because metal-rich ex-situ clusters from massive satellites have been accreted onto the central halo.

The dependence on the mass threshold is significant. In the full sample used in this work, $\log (M_{\rm lim}/\Msun) = 5.0$, the in-situ and ex-situ CDFs extend to much lower metallicities than distributions at mass thresholds corresponding to observational limits. A higher mass threshold shifts the distribution to higher metallicities. This implies a mass-metallicity relation, sometimes referred to as blue tilt. The in-situ fraction of the population also increases with a higher mass threshold. The in-situ fraction combined from all three model realizations increases from $20\%$ to $23\%$ to $27\%$ to $33\%$ for $\log (M_{\rm lim}/\Msun) = 5.0$, 6.0, 6.3, and 6.5, respectively.

\subsection{Mass-Metallicity Relation}

\begin{figure}
    \centering
    \includegraphics[width=\linewidth]{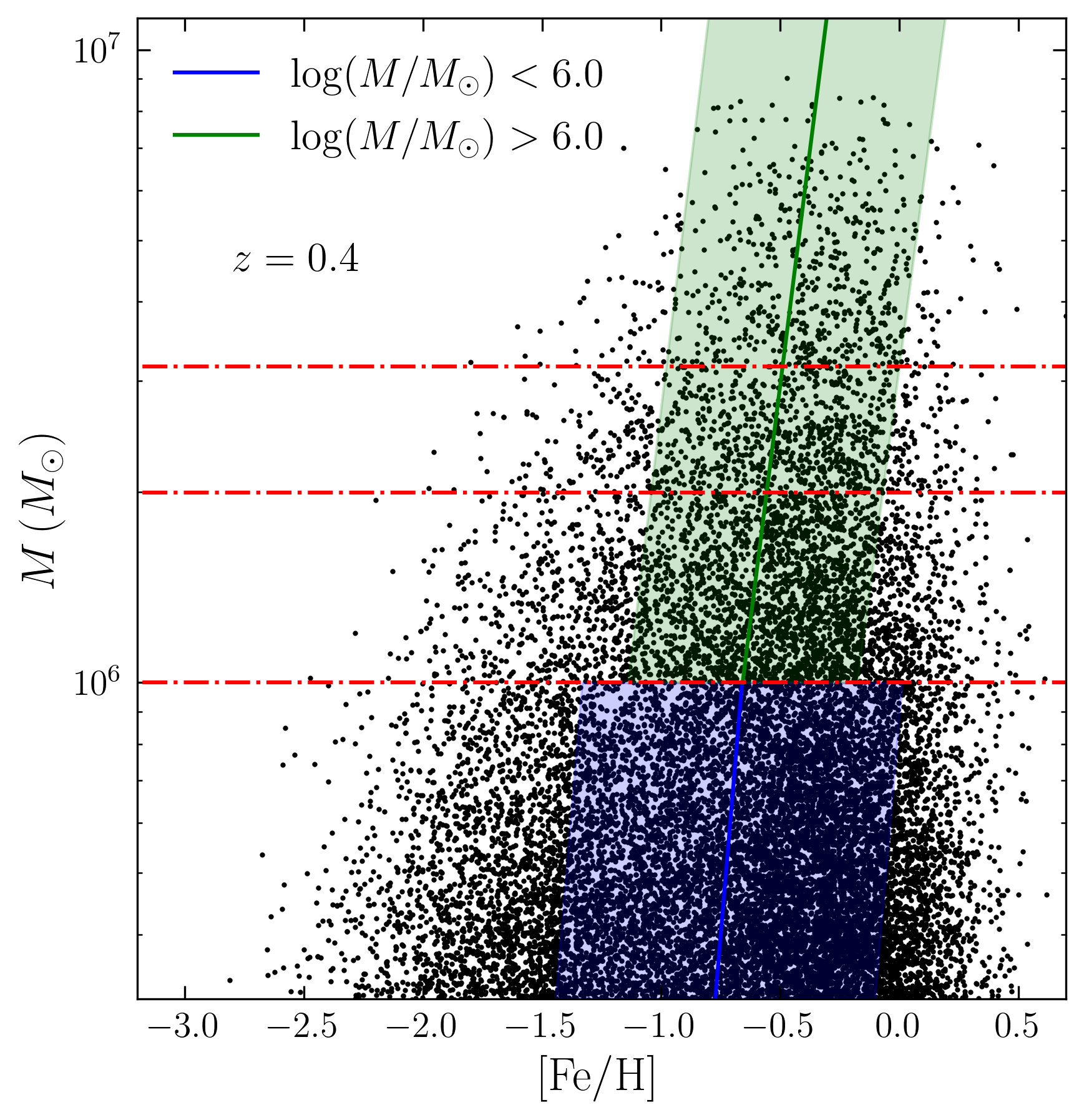}
    \caption{Mass-metallicity relation for the stacked data set of all model realizations at $z = 0.4$, for $R_{\rm lim} = 1$~Mpc. We fit a separate relation above (green) and below (blue) $M=10^6 \Msun$. The shaded regions represent the intrinsic scatter and uncertainty of the best fit. The horizontal red lines illustrate typical observational mass limits $\log (M/\Msun) = 6.0$, 6.3, and 6.5.}
    \label{fig:mass-metallicity}
\end{figure}

We fit the mass-metallicity relation using the combined distribution from all three model realizations. We adopt a functional form
\begin{equation}
    [\mathrm{Fe/H}] = \alpha \log_{10} \frac{M}{10^6 \Msun} +C +\epsilon.
    \label{eq: mass-metal relation}
\end{equation}
Here, $\alpha$ and $C$ are the slope and intercept parameters and $\epsilon$ represents the intrinsic scatter of each data point described by the normal distribution ${\cal N}(0,\sigma^2)$. We set a pivot point at $10^6 \Msun$. We expect the mass-metallicity relationship to be only a property of high-mass GCs, so we performed the fit both above and below the pivot mass. To fit this functional form, we used the likelihood function
\begin{equation}
    {\cal L} = \prod_i {1 \over \sqrt{2\pi} \sigma} \exp\left[-{(\feh - C - \alpha \log (M/10^6 \Msun))^2 \over 2 \sigma^2}\right].
    \label{mass-metal likelihood}
\end{equation}
$\sigma$ is the standard deviation of the normal distribution that represents the intrinsic scatter. We minimized $-\ln {\cal L}$ with \textsc{Scipy} \texttt{minimize} to obtain the best-fit values for $\alpha$, $C$ and $\sigma$. We preformed a traditional bootstrap to estimate uncertainties with 1000 realizations.

We fit the relation at all 6 epochs above and below the pivot mass. Below the pivot mass, we still applied a lower mass limit of $\log (M_{\rm lim}/\Msun) = 5.0$. The average slope in the high-mass regime across all epochs is $\alpha = 0.34 \pm 0.02$, where the uncertainty is the standard deviation between epochs. The average single-epoch bootstrap uncertainty is slightly higher $\bar{\sigma}_{\rm boot}=0.03$. The slope  varies from $\alpha = 0.29 \pm 0.03$ to $\alpha = 0.34 \pm 0.03$ to $\alpha = 0.35 \pm 0.03$ for $z = 1$, 0.4, and 0, respectively. Below the pivot mass, the relation is weaker, with the average slope across all epochs is $\alpha = 0.22 \pm 0.02$ with $\bar{\sigma}_{\rm boot}=0.02$. It varies somewhat with time from $\alpha = 0.187 \pm 0.01$ to $\alpha = 0.225 \pm 0.01$ to $\alpha = 0.222 \pm 0.02$ for $z = 1$, 0.4, and 0, respectively. Figure~\ref{fig:mass-metallicity} shows the mass-metallicity distribution at $z = 0.4$ with the best-fit relations above and below the pivot mass.

The best-fit slope changes with a higher limiting mass. Considering a single epoch $z = 0.4$, the slope varies from $\alpha = 0.33 \pm 0.03$ to $\alpha = 0.46\pm0.06$ to $\alpha = 0.4 \pm 0.1$ for $\log (M_{\rm lim}/\Msun) = 6.0$, 6.3, and 6.5, respectively. At higher mass thresholds, the slope is poorly determined because of an insufficient number of model GCs.

To investigate further, we split the cluster distribution into its in-situ and ex-situ components and fit the mass-metallicity relation for each component separately. For the in-situ clusters, the average slope across 6 epochs is marginal both above and below the pivot mass. The mean slope is $\alpha = -0.08 \pm0.01$ ($\bar{\sigma}_{\rm boot}=0.04$) above the pivot mass and $\alpha = -0.02 \pm0.01$ ($\bar{\sigma}_{\rm boot}=0.01$) below the pivot mass. For the ex-situ clusters, the average slope is higher than for the combined population in both mass regimes. The average slope is $\alpha = 0.42 \pm0.02$ ($\bar{\sigma}_{\rm boot}=0.04$) above the pivot mass and $\alpha = 0.28 \pm 0.02$ ($\bar{\sigma}_{\rm boot}=0.02$) below the pivot mass.

Similarly, we divided the distribution into metal-rich and metal-poor components using a simple metallicity threshold $\feh=-0.55$, discussed later in \S\ref{sec:metallicity_classification}. Both above and below the pivot mass, the average slope of metal-rich GCs is close to zero. The mean slope is $\alpha = -0.03 \pm0.02$ ($\bar{\sigma}_{\rm boot}=0.02$) above the pivot mass and $\alpha = -0.02 \pm0.01$ ($\bar{\sigma}_{\rm boot}=0.01$) below the pivot mass. Similar to ex-situ GCs, the metal-poor GCs have a higher slope than the combined populations. The mean slope is $\alpha = 0.37 \pm0.02$ ($\bar{\sigma}_{\rm boot}=0.04$) above the pivot mass and $\alpha = 0.31 \pm0.02$ ($\bar{\sigma}_{\rm boot}=0.01$) below the pivot mass. 

The mass-metallicity relationship is thought to be a property of metal-poor GCs at the high-mass end of the GC mass function. It has been observed in nearby galaxies massive enough to populate the high-mass end \citep[e.g.,][]{Harris2006, Mieske2006, Strader2006, Harris2017b}. Observations have found that the slope of the relationship varies significantly between galaxies, with some showing no trend at all. No similar relationship has been observed for metal-rich GCs.

\begin{figure*}
    \centering
    \includegraphics[width=\linewidth]{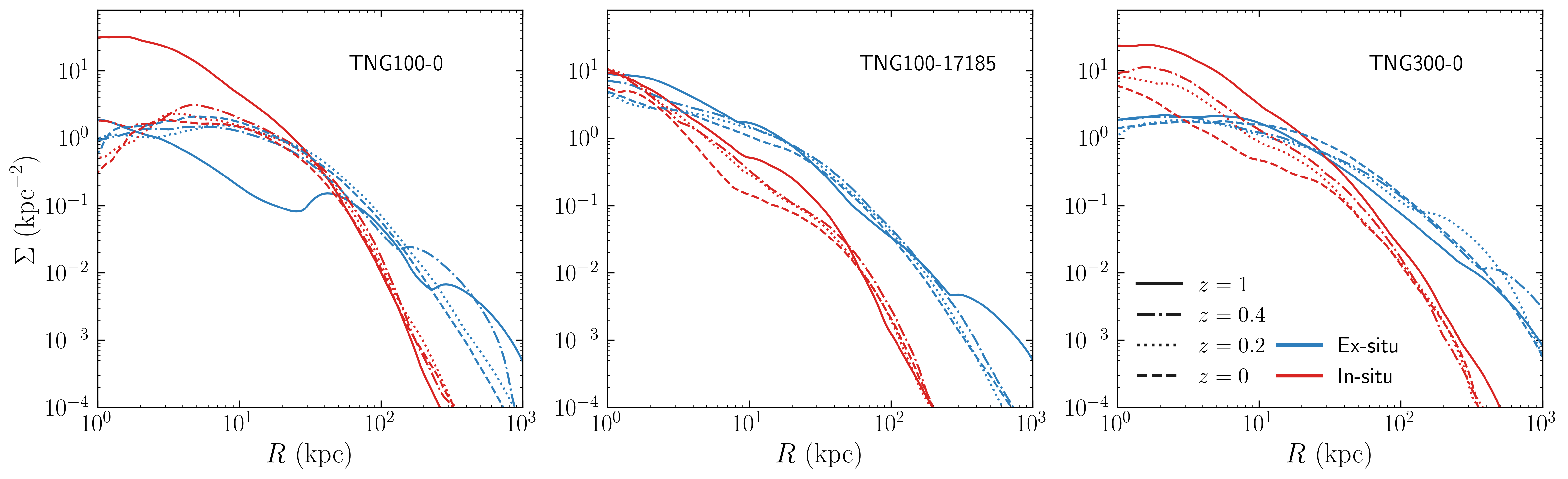}
    \caption{In-situ and ex-situ surface density profiles for each model realization at 4 selected epochs of evolution, for $\log(M_{\rm lim}/\Msun) =5.0$. Ex-situ clusters dominate at large radii and in-situ clusters dominate in the core of the BCG.}
    \label{fig: in-situ/ex-situ surf den}
\end{figure*}

\cite{Choksi2018} investigated this relationship in a previous version of the GC model. They showed that it arises because metal-poor GCs form preferentially in low-mass halos ($M_h \lesssim 10^{11} \Msun$) at early times ($z \gtrsim 3$). The total amount of cold gas available to form clusters is low, so the high-end of the ICMF is not fully sampled. Thus, high-mass metal-poor GCs are only formed in massive galaxies with large reservoirs of cold gas. The gas mass scales with the stellar mass, and the stellar mass scales with galaxy metallicity. GCs inherit the metallicity of their formation environment; hence, most massive metal-poor GCs tend to be metal-enhanced relative to less massive GCs. This is not observed for metal-rich GCs because they form in massive host galaxies where there is a sufficient amount of gas to fully sample the ICMF. This also explains why the mass-metallicity relation is insignificant for in-situ clusters but relevant for ex-situ clusters. Figure~\ref{fig:metallicity_cdfs} shows that in-situ clusters in our model are always metal-rich ($\feh \gtrsim -1.5$). Thus, in-situ clusters form in the giant galaxies where the ICMF is fully sampled, and the mass-metallicity relation does not arise. Conversely, ex-situ clusters can be metal-poor and form under the conditions leading to the mass-metallicity trend.

\section{In-situ and Ex-situ Globular Clusters}
\label{sec:in-situ_ex-situ}

The contributions of in-situ and ex-situ clusters are an important indicator of the formation history of the GC population and the host galaxy. We investigated the distribution of model in-situ and ex-situ clusters around their hosts and tested potential observational methods for distinguishing the two populations. 

\subsection{The Crossover Radius}

Figure~\ref{fig: in-situ/ex-situ surf den} shows the model surface density profiles of in-situ and ex-situ GCs. In general, in-situ clusters dominate near the BCG core, and ex-situ clusters dominate at large radii. The radius where the local population switches from in-situ to ex-situ is the intersection of the surface density profiles. This radius varies significantly between the different model realizations and evolves over time. We call the radius where the population shifts from in-situ dominated to ex-situ dominated the crossover radius $R_{\rm cross}$. We use the crossover radius as a quantitative metric for analyzing the distribution of in-situ and ex-situ GCs.

\begin{figure}
    \centering
    \includegraphics[width=\linewidth]{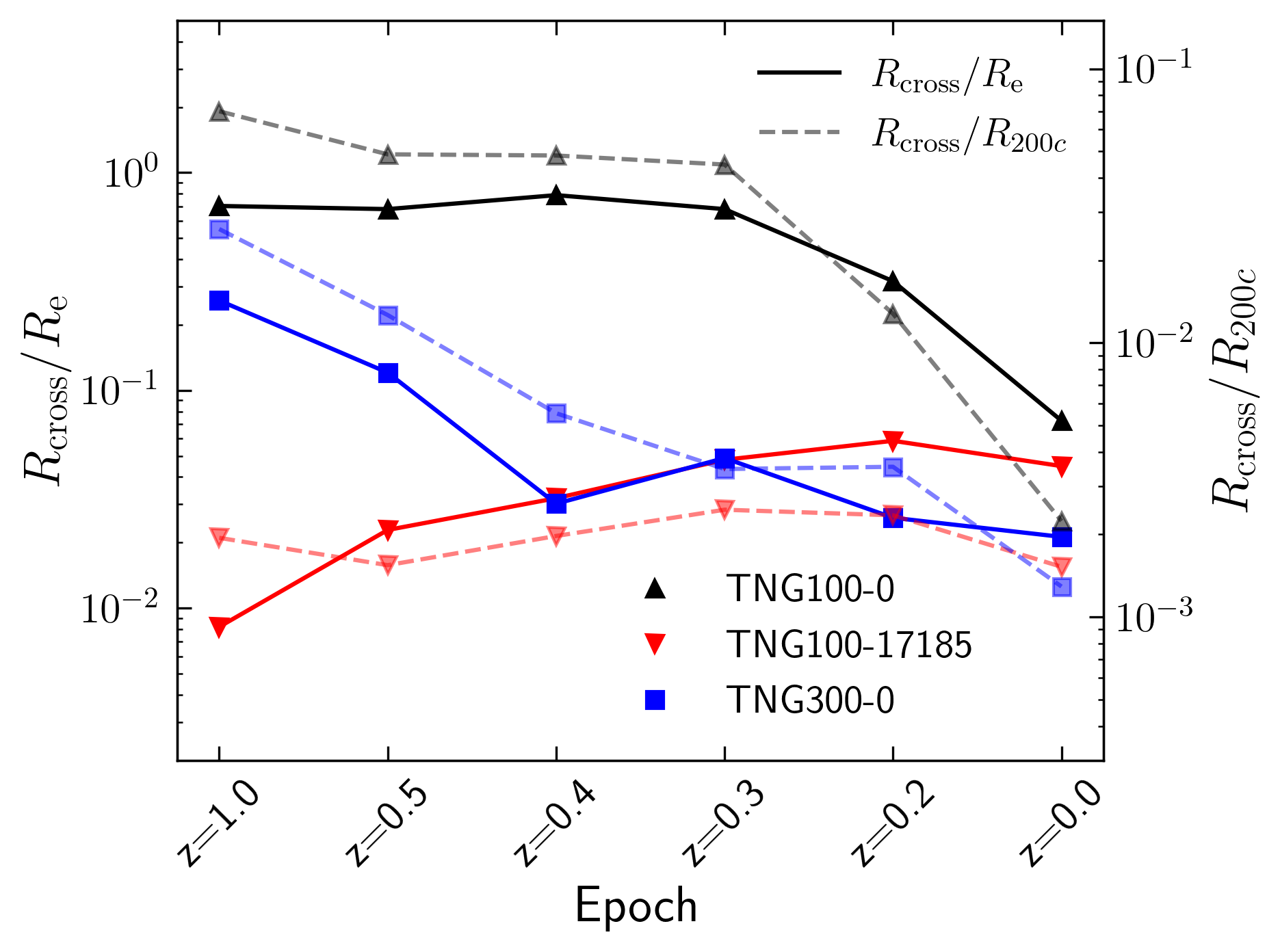}
    \caption{The crossover radius $R_{\rm cross}$ scaled by the half-mass radius of the GC system $R_{\rm e}$ (left axis) and by the virial radius of the host halo $R_{200c}$ (right axis).}
    \label{fig:crossoverRscaled}
\end{figure}

We located $R_{\rm cross}$ by modeling the surface density profile KDEs with linear interpolating functions. We fixed a high minimum KDE bandwidth $h_{\rm min} = 0.2$ to ensure a smooth profile when calculating the interpolating function. We apply a limiting mass $\log(M_{\rm lim}/\Msun) =5.0$. The radius where the in-situ and ex-situ interpolating functions intersect is the crossover radius. The values of $R_{\rm cross}$ are reported in Table~\ref{tab:crossover_r}.

Figure~\ref{fig:crossoverRscaled} shows the crossover radii scaled by the half-mass radius of the host GC population $R_{\rm e}$ and by the virial radius $R_{200c}$ of the host galaxy. As discussed in \S\ref{sec:satellites}, the half-mass radius is the median 2D projected radius of GCs from the host galaxy which we measured from the GC catalogs with $\log(M_{\rm lim}/\Msun) = 5.0$ and $R_{\rm lim} = 1$~Mpc. The virial radius is defined by an overdensity 200 times the critical density of the universe at the relevant epoch. We used $R_{200c}$ values reported in the TNG group catalogs for each model realization at the snapshots corresponding to the 6 epochs investigated in this work. TNG300-0 is a larger host galaxy; therefore, we expect it to have larger crossover radii. Scaling by $R_{\rm e}$ and $R_{200c}$ allows us to compare the crossover radii between host galaxies of different sizes. 

Despite having similar masses, TNG100-0 and TNG100-17185 have dramatically different crossover radii. This is reflected in their surface density profiles. Figure~\ref{fig: in-situ/ex-situ surf den} shows that TNG100-0 has similar in-situ and ex-situ surface densities between $5-50$ kpc at $z<1$, and the crossover is located around 50~kpc. TNG100-17185 instead shows nearly complete ex-situ dominance.

\begin{table*}
\centering
\caption{Host galaxy mass, half-mass radius $R_{\rm e}$, crossover radius $R_{\rm cross}$, and metallicity crossover radius $R_{\rm cross,m}$ over time.}
\label{tab:crossover_r}
\begin{tabular}{ccccccccccccc}
\toprule
  & \multicolumn{4}{c}{TNG100-0} & \multicolumn{4}{c}{TNG100-17185} & \multicolumn{4}{c}{TNG300-0} \\
  \cmidrule(lr){2-5} \cmidrule(lr){6-9} \cmidrule(lr){10-13}
 z & $\log_{10}{M_{200c}}$ & $R_{\rm e}$ & $R_{\rm cross}$ & $R_{\text{cross,m}}$  & $\log_{10}{M_{200c} }$ & $R_{\rm e}$ & $R_{\rm cross}$ & $R_{\text{cross,m}}$  & $\log_{10}{M_{200c}}$ & $R_{\rm e}$ & $R_{\rm cross}$  & $R_{\text{cross,m}}$\\
   & $(\Msun)$ & (kpc) & (kpc) & (kpc) &  $(\Msun)$ & (kpc) & (kpc) & (kpc) & $(\Msun)$ & (kpc) & (kpc) & (kpc) \\
\midrule
0.0 & 14.58 & 47.0 & 3.4 & 100.5 & 14.58 &  51.9 & 2.3 & 20.6 & 15.19 & 147.9 & 3.1 & 16.0 \\
0.2 & 14.41 & 50.2 & 16.0 & 102.0 & 14.34 & 47.4 & 2.8 & 29.9 & 15.09 & 285.7 & 7.4 & 14.6 \\
0.3 & 14.37 & 76.7 & 52.2 & 156.4 & 14.14 & 50.2 & 2.4 & 27.2 & 15.12 & 147.9 & 7.2 & 18.4 \\
0.4 & 14.37 & 68.7 & 54.2 & 89.9 & 14.08 &  55.9 & 1.8 & 30.6 & 15.14 & 373.2 & 11.3 & 22.9 \\
0.5 & 14.35 & 76.6 & 52.1 & 92.4 & 14.00 &  55.2 & 1.3 & 31.7 & 15.09 & 197.7 & 23.9 & 30.3 \\
1.0 & 14.09 & 71.3 & 50.1 & 73.6 & 13.91 & 147.9 & 1.2 & 76.4 & 14.71 & 115.6 & 29.9 & 37.9 \\
\bottomrule
\end{tabular}
\end{table*}

\begin{figure}
    \centering
    \includegraphics[width=\linewidth]{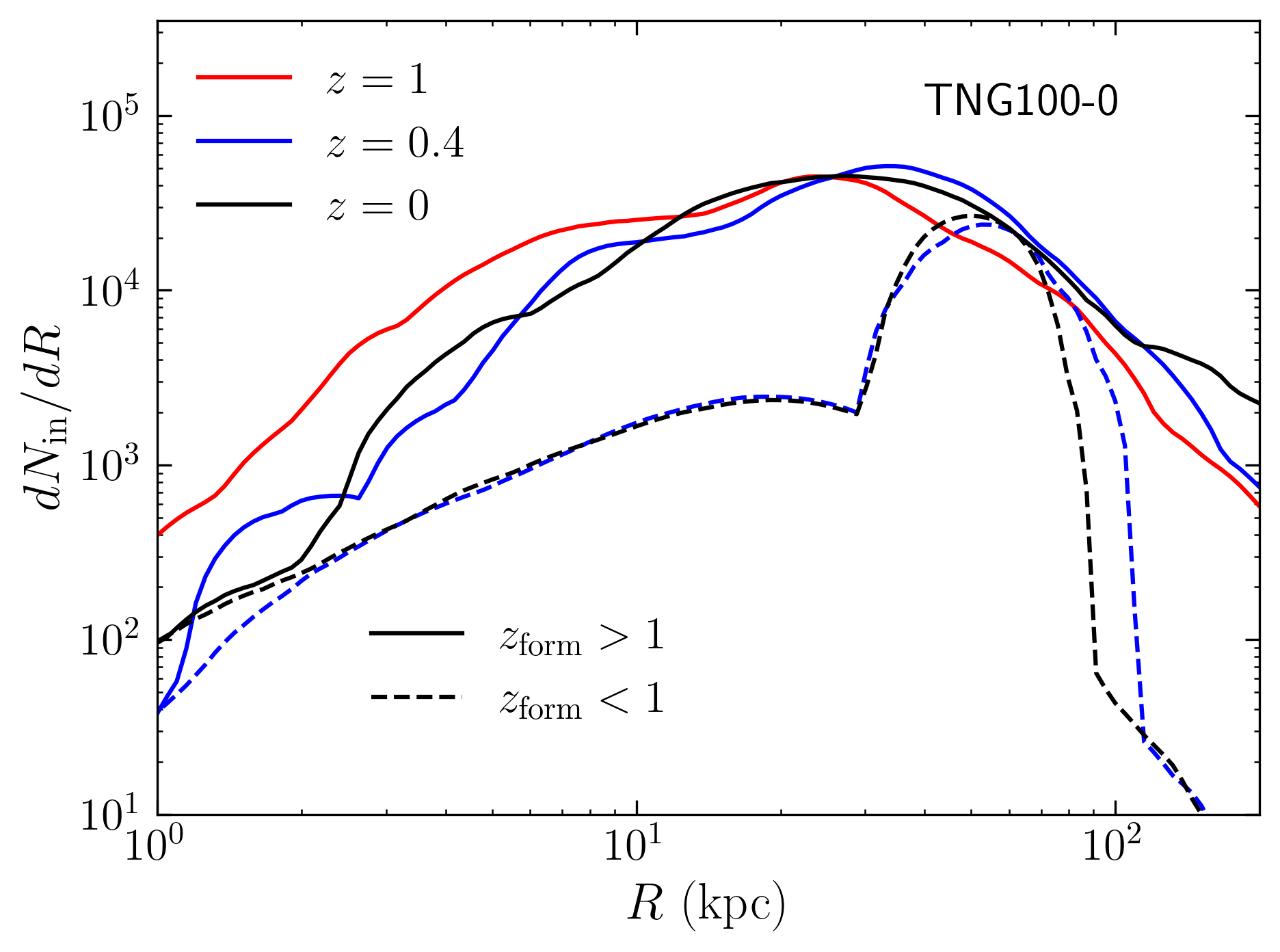}
    \caption{The radial distributions of in-situ GCs in TNG100-0 formed before (solid) and after $z=1$ (dashed) at three labeled epochs. There is a noticeable decrease in the number of old in-situ GCs within 10 kpc between $z = 1$ and later epochs. This decrease is caused by a spatial redistribution of GCs following a major merger.}
    \label{fig:merger_investigation}
\end{figure}

The extended in-situ dominated region in TNG100-0 is explained by a late merger that occurred just after $z=1$. The merger perturbed the galactic potential, resulting in a mixing of both populations. We investigated the impact of the merger by plotting the radial distribution of in-situ GCs in TNG100-0 formed before $z=1$ and after $z=1$ in Figure~\ref{fig:merger_investigation}. There are three processes that could alter the GC distribution: formation, disruption, and displacement. By isolating clusters formed after $z=1.0$, we can determine the relative importance of cluster formation. Figure~\ref{fig:merger_investigation} shows a young cluster peak around 50~kpc indicating continued cluster formation in this region. This peak is below the radial distribution for old clusters at the same radii, indicating that its contribution to the total in-situ distribution is small. From the distribution of clusters formed before $z = 1$, we see a significant decrease in the number of clusters within 10~kpc between $z = 1$ and later epochs. The decrease cannot be explained only by disruption because the disruption rate is relatively constant with time, and more time passes between the epochs of $z = 0.4$ and $z = 0$ (4.3~Gyr) than between $z = 1$ and $z = 0.4$ (3.5~Gyr), yet the radial distribution is stable after $z = 0.4$. Thus, the shift at $z<1$ is best explained as a redistribution of in-situ clusters around the galaxy that occurred during the major merger event. This redistribution produces the mixed dominance region in the surface density profile and a larger crossover radius. 

It is also relevant that TNG100-0 and TNG100-17185 have very different in-situ fractions at $z = 1$: 45\% for TNG100-0 and 11\% for TNG100-17185. A larger initial in-situ fraction will naturally produce a larger crossover radius by extending the region of its dominance. We expect that the difference in the in-situ fraction of these two halos is a product of their unique formation histories. 

We tested how a moderate change and a significant change in the mass threshold affect the crossover radius. First, we recalculated the crossover radii with a limiting mass $M_{\rm lim} = 2\times10^{5} \Msun$. This has little effect. The crossover radius decreased by 0.22--4.7 kpc (16th--84th percentile range). Then we recalculated the crossover radius for a typical observable mass limit $\log (M_{\rm lim}/\Msun) =6.3$, and the crossover disappeared for TNG100-17185. The distribution is entirely ex-situ dominated. For the other two model realizations, the median shift in the crossover radius is only $0.1$~kpc but with considerable variation. The crossover radius changed by $-16$~kpc to $+4.9$~kpc in the 16th--84th percentile range. At higher mass thresholds corresponding to observational limits, the model GC populations are too small to accurately measure the crossover radius.

\begin{figure*}
    \centering
    \includegraphics[width=\linewidth]{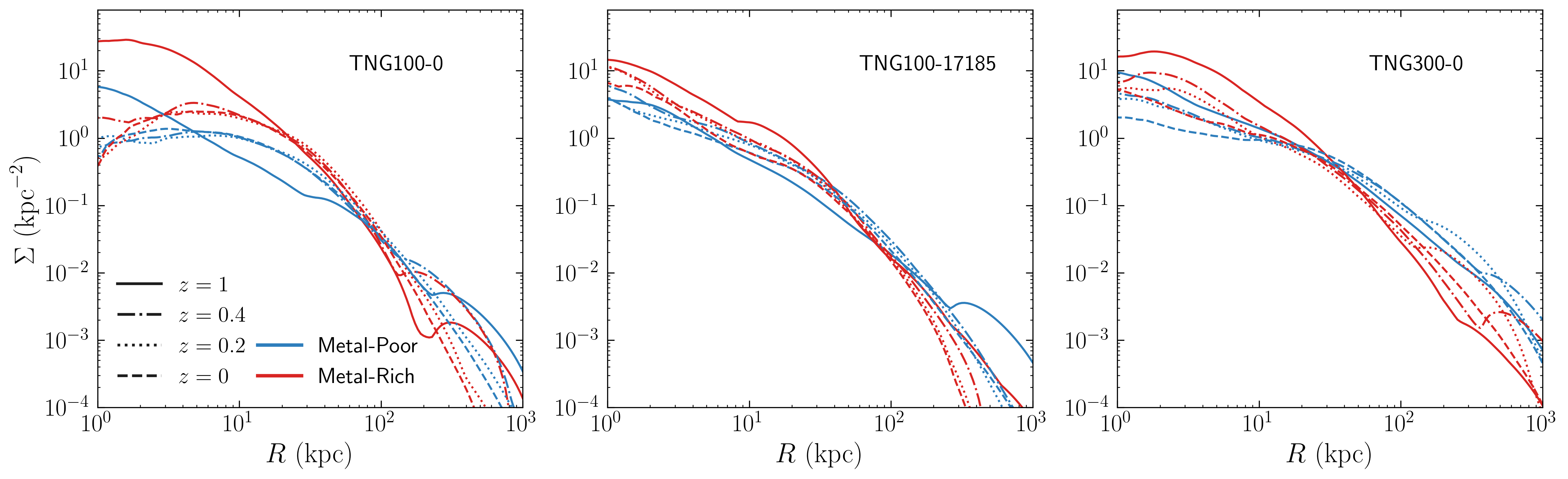}
    \caption{Metal-poor and metal-rich surface density profiles, split at $\feh=-0.55$, for each model realization with $\log(M_{\rm lim}/\Msun) =5.0$. The crossover radius from metal-rich to metal-poor clusters occurs at larger $R$ than the crossover from in-situ to ex-situ clusters.}
    \label{fig:surf_den_metal}
\end{figure*}

\subsection{Metallicity Classification}
\label{sec:metallicity_classification}

We tested using metallicity as a proxy for in-situ and ex-situ cluster classification, because it is commonly applied for observations. To determine the optimal metallicity threshold for sorting GCs as in-situ and ex-situ, we maximized balanced accuracy $A_B$, which is defined as
\begin{equation}
    A_B = \frac{1}{2} \left( \frac{TP}{TP+FN} + \frac{TN}{TN+FP}\right)
    \label{eq: balanced accuracy}
\end{equation}
where $TP, \ TN,\ FP,$ and $FN$ are, respectively, the number of True Positives, True Negatives, False Negatives, and False Positives. In data sets where one class dominates, balanced metrics like $A_B$ are more reliable for model performance than traditional accuracy. The model could predict that all objects belong to the dominant class and still have a high traditional accuracy. Balanced metrics avoid this problem by giving equal weight to the performance of both classes. In general, ex-situ clusters are dominant, so we used the balanced accuracy to ensure that our threshold is correctly discriminating between the two classes.  

We determined the threshold that maximizes $A_B$ for each model realization and epoch, using the limiting mass and radius $\log (M_{\rm lim}/\Msun) =5.0$ and $R_{\rm lim} = 1$~Mpc. By averaging the optimal thresholds of all model realizations and epochs, we determined a universal threshold of $[\mathrm{Fe/H}] = -0.55$. Applying this threshold to all model realizations and epochs gives a balanced accuracy $A_B = 0.785$, and a traditional accuracy $A = 0.716$.

\begin{figure}
    \centering
    \includegraphics[width=\linewidth]{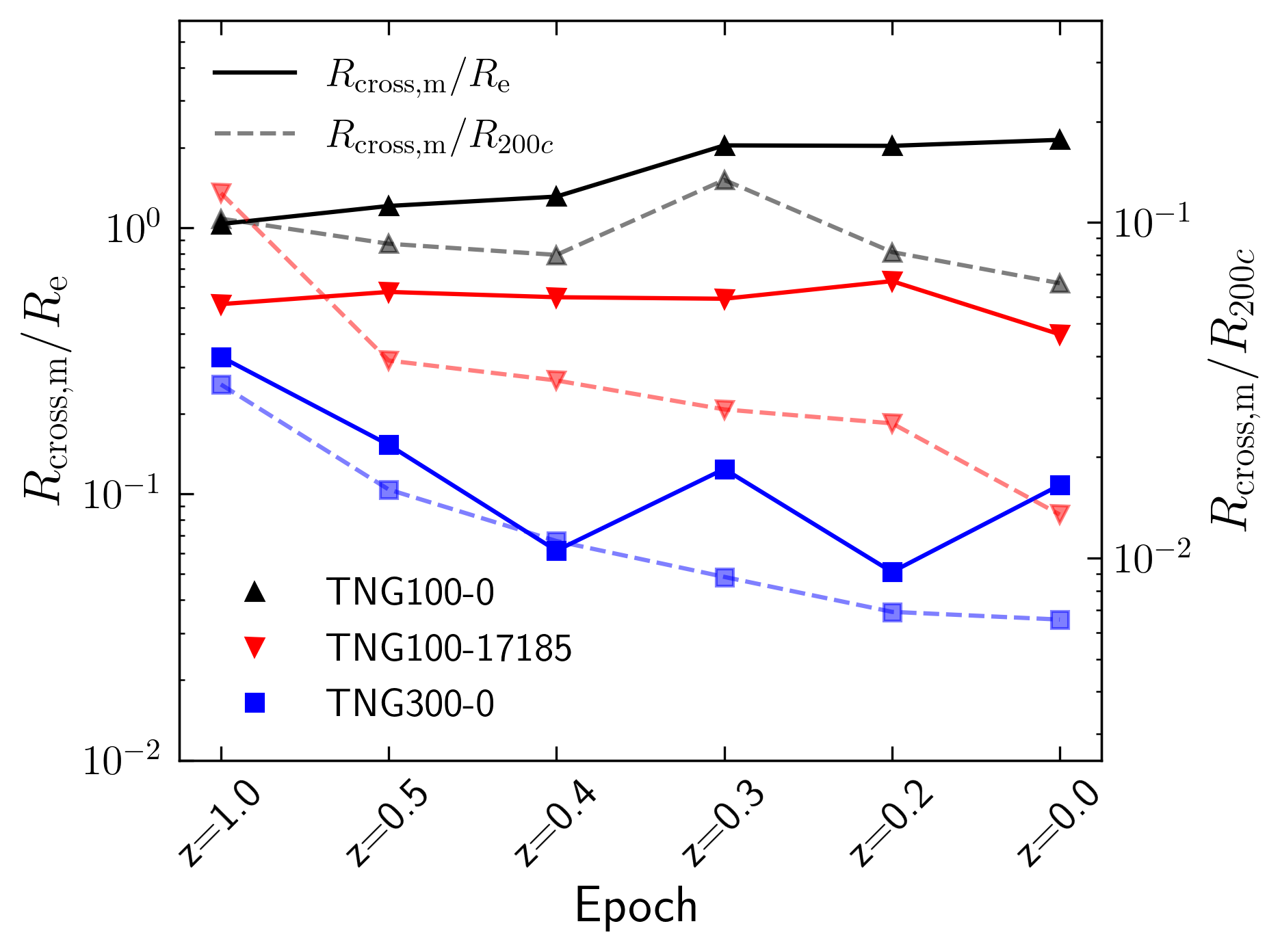}
    \caption{Crossover radius based on metallicity threshold $R_{\rm cross,m}$ scaled by the half-mass radius of the GC system $R_{\rm e}$ (left axis) and by the virial radius of the host halo $R_{200c}$.}
    \label{fig:crossover_R_metal}
\end{figure}

We divided GCs into metal-poor $[\mathrm{Fe/H}] < -0.55$ and metal-rich $[\mathrm{Fe/H}] > -0.55$ populations. Figure~\ref{fig:surf_den_metal} shows the surface density profiles of metal-poor and metal-rich populations. We computed the metallicity-based crossover radius $R_{\rm cross,m}$ following the same procedure as for in-situ vs. ex-situ clusters. $R_{\rm cross,m}$ is on average larger than $R_{\rm cross}$ by 10--78~kpc (16th--84th percentile range). Figure~\ref{fig:crossover_R_metal} shows the metallicity crossover radius $R_{\rm cross,m}$ scaled by the half-mass radius $R_{\rm e}$ and by the virial radius $R_{200c}$. Table~\ref{tab:crossover_r} reports the mass of the host galaxy, the half-mass radius $R_{\rm e}$, the crossover radii $R_{\rm cross}$, and the metallicity crossover radii $R_{\rm cross,m}$ for easy comparison.

\subsection{Machine Learning Classification}
\label{sec:ML_classification}

We trained supervised machine learning (ML) classifiers to distinguish in-situ and ex-situ clusters at the 6 epochs investigated by the model and with various limiting masses and radii. The advantage of ML methods is that they can simultaneously leverage all observable features to classify clusters and are not restricted to using linear decision boundaries. The mass thresholds we explored were $\log (M_{\rm lim}/\Msun) = 5.0$, 6.0, and 6.3, and the radial cutoffs were $R_{\rm lim} = 100$, 200, and 300~kpc. The classifiers were trained on the three model features: mass, metallicity, and 2D projected distance. We optimized over different ML methods and their hyperparameters to find a classifier with the best balanced accuracy. Like for the metallicity threshold, we chose a balanced metric to give equal weight to the performance of both classes. 

We divided the data into training, validation, and testing data sets. We ensured that the training set is evenly composed of data from each model realization, using the following procedure. We found the size of the data set for the smallest halo ($N_{\rm min}$) and then randomly sampled $0.7N_{\rm min}$ GCs from each model realization to build the training set. For validation, we randomly sampled $0.15N_{\rm min}$ from each model realization to create 3 individual validation sets for each halo and 1 combined validation set. Any unseen GCs left over after training and validation were used in the test set.

To find the best classifier for each redshift, mass threshold, and radial threshold we tested 5 different \textsc{Scikit-learn} methods: Histogram Gradient Boosting Classifier, Random Forest Classifier, K Nearest Neighbors Classifier, Logistic Regression Classifier, and Support Vector Machines Classifier. For the Support Vector Machines Classifier, we scaled the data set to have a mean of 0 and a standard deviation of 1, using \textsc{Scikit-learn} \texttt{StandardScaler}. We optimized the hyperparameters of the models using \textsc{Scikit-learn}'s \texttt{RandomizedSearchCV} function. For each classifier, we pass \texttt{RandomizedSearchCV} a grid of possible hyperparameters. The function randomly samples the hyperparameter grid 100 times to find the set of hyperparameters that maximizes the balanced accuracy. It estimates the balanced accuracy using $K$-fold cross-validation with $K = 5$. \texttt{RandomizedSearchCV} runs significantly faster than exhaustive search routines like \texttt{GridSeachCV}. Increasing the number of samples that \texttt{RandomizedSearchCV} takes from the hyperparameter grid does not significantly improve the balanced accuracy. Thus, \texttt{RandomizedSearchCV} effectively maximizes the balanced accuracy without requiring a computationally intensive exhaustive search of the hyperparameter space. After \texttt{RandomizedSearchCV} identifies the best hyperparameters for the classifier, we test it on the validation dataset. We find the balanced accuracy for the combined validation set and the validation set for each halo. We take the average of these four balanced accuracies and search for the classifier with the highest average. We determined the best classifiers and ran these on the test data to obtain the final performance metrics. 

\begin{figure}
    \centering
    \includegraphics[width=\linewidth]{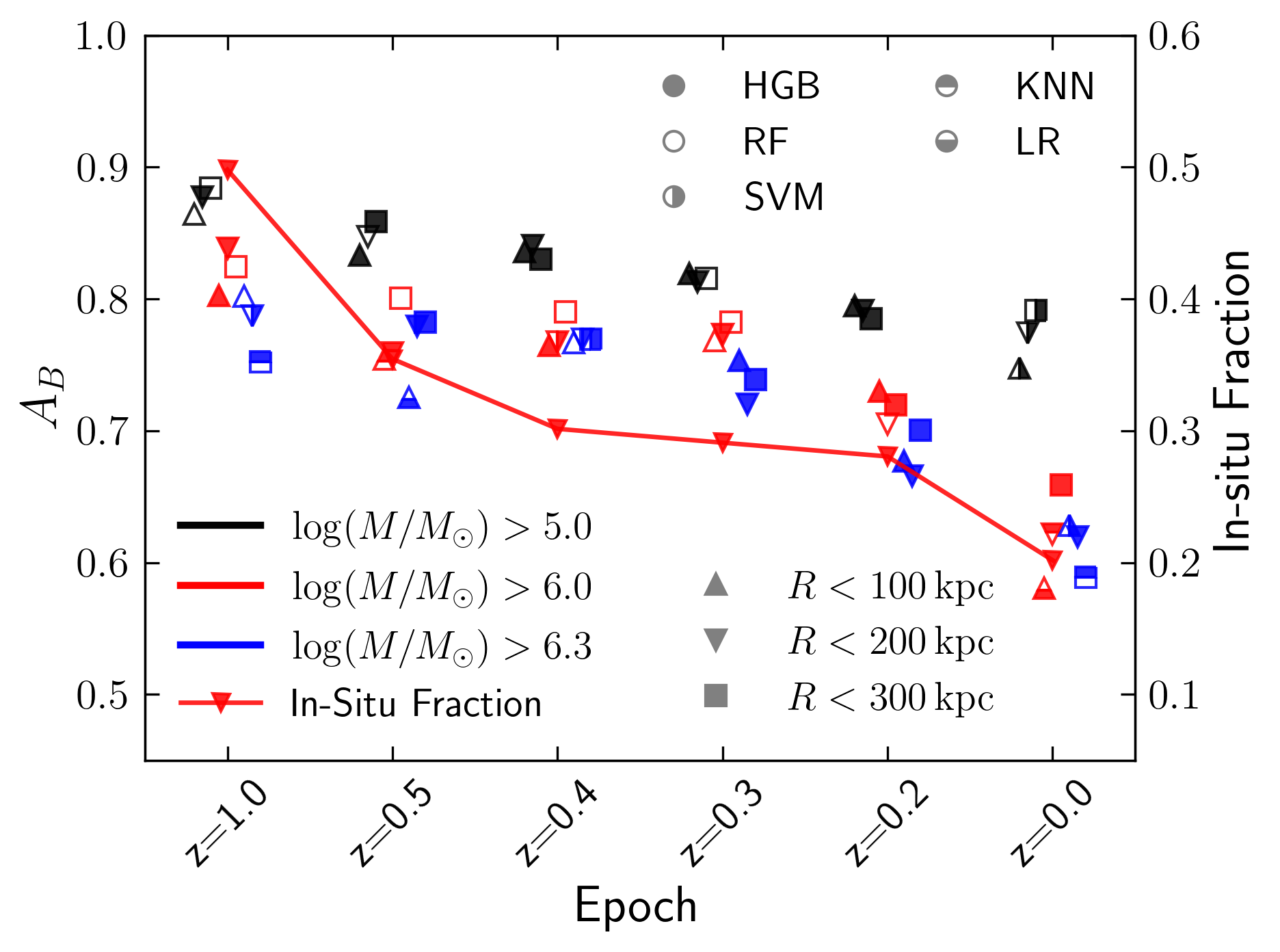}
    \caption{Balanced accuracy of the best ML classifiers on the test set (left axis). Each color represents a different mass threshold and each symbol represents a different radial threshold. The filling of the symbol corresponds to the best classifier for that data set. The solid line is the in-situ fraction of the test set with mass threshold $\log(M_{\rm lim}/\Msun) = 6.0$ and radial threshold $R_{\rm lim} = 200$~kpc (right axis). The balanced accuracy of the classifiers decreases with redshift, similarly but less strongly than the decrease in the in-situ fraction.}
    \label{fig: balance_accuracy}
\end{figure}

We found that the optimal classifiers are predominately Histogram Gradient Boosting Classifiers and Random Forest Classifiers, with a few data sets preferring Support Vector Machines, K Nearest Neighbors, and Logistic Regression. Figure~\ref{fig: balance_accuracy} shows the balanced accuracy of the classifiers on the test set. The overplotted line shows the evolution of the in-situ fraction of the test data. Both decrease with redshift. As the size of the smaller class decreases, it becomes more difficult to correctly classify. However, the balanced accuracy does not drop as significantly as the in-situ fraction, indicating that we obtained better performance than expected from the in-situ fraction alone. For the full sample used in this work, the average balanced accuracy and standard deviation is $A_{B} = 0.82\pm0.04$. At observational limits it drops to $A_{B} = 0.75\pm0.07$ and $A_{B} = 0.72\pm0.07$ for $\log(M_{\rm lim}/ \Msun) = 6.0$ and 6.3, respectively. The cause of the drop is likely the size of the training set. The average size of the training sample decreases from $17,000$ to $2,000$ between $\log(M_{\rm lim}/\Msun) = 5.0$ and 6.0, and further to $700$ for $\log(M_{\rm lim}/\Msun) = 6.3$. The average spread due to the choice of radial threshold is smaller, only $\sim$2\%. 
The traditional accuracy does not equally measure the performance of both classes and is thus largely insensitive to the decrease in the in-situ fraction. It is generally consistent across epochs. It is less sensitive to the choice of limiting mass and thus the size of the training set. It varies from $A = 0.84 \pm 0.07$ to $A = 0.81 \pm 0.03$ to $A = 0.78 \pm 0.03$ for $\log(M_{\rm lim}/\Msun) = 5.0$, 6.0, and 6.3.

If we optimize traditional accuracy instead of balanced accuracy, we obtain similar traditional accuracy scores, but the balanced accuracy drops more significantly with time as the in-situ fraction decreases. Interestingly, if we use a test set composed evenly from each model realization, the balanced accuracy is slightly worse. This is because the large test dataset is composed mostly of GC from TNG300-0 and TNG100-0, because TNG-17185 has the smallest GC population by number. TNG300-0 and TNG100-0 have higher in-situ fractions (28\% and 45\% at $z=1.0$, respectively) than TNG-17185 (11\% at $z=1$) so the large test set has a higher in-situ fraction than the even test set. The performance of the classifiers on the in-situ class is better in the large test set because the in-situ fraction is higher.

\section{Discussion and Comparison with Observations} 
\label{sec:discussion}

The goal of our analysis is to understand the globular cluster populations of BCGs in distant galaxy clusters. There are important limitations of our model for its use in this investigation. 1) The model parameters were calibrated with observation of GCs of Milky-Way-like galaxies, and the model has been validated only on Milky-Way-sized halos. In this work, we assume that the model extends to giant elliptical galaxies. 2) Our model for cluster formation draws from the ICMF with an exponential cutoff at $10^{7} \Msun$. Observations of giant BCGs detect significant GC populations above this mass. Future work can explore potential adjustments to the cutoff mass in giant galaxy systems. 3) We only study the GC systems around three simulated BCG-like galaxies. Three halos is not sufficient to represent the full diversity of BCGs. Additionally, the TNG100 halos are less massive than most of the BCG systems where GCs have been observed. While the mass of the TNG300 halo is on par with observed BCGs, the simulation resolution is coarser than we would like. With these caveats in mind, what follows is a discussion of the interpretation of our findings and a comparison with previous work. 

\subsection{Mass Function}
\label{sec:mass_func_discussion}

Our model predicts that the GC mass function of giant galaxies gets shallower over time. The slope is closest to 2 at earlier epochs before flattening to about 1.8--1.9 at $z = 0$ due to preferential disruption of low mass clusters. The expected value $\beta \approx 2$ is inferred from the ICMF, thus at earlier times, before significant disruption has occurred, the mass function will be closer to the initial mass function. This is recovered by our model. 


Most observational studies of the luminosity function (LF) of BCGs fit the profile with a log-normal distribution of the form
\begin{equation}
    \frac{dN}{d\log_{10}M} \propto\exp{\left(-\frac{(\log M - \log M_{\rm TO})^2}{2 \sigma_{\log M}^2}\right)},
\end{equation}
where $\log M_{\rm TO}$ and $\sigma_{\log M}$ are the mean and standard deviation in $\log M$ space. This can be compared with our measurement using an effective power law slope for the log-normal defined as
\begin{equation}
   \beta_{\rm eff}(M) = -\frac{\ln(dN/dM)}{d\ln M} = 1 +\frac{\log M - \log M_{\rm TO}}{\sigma^2_{\log M}\ln{10}}.
   \label{eq:eff_powerlaw}
\end{equation}
The standard deviation relates to the dispersion in magnitude space as $\sigma_{\log M} = 0.4\,\sigma_m$. The turnover mass $M_{\rm TO}$ is calculated from the observed turnover magnitude $m_0$. Equation~5 of \cite{2024Harris} gives a relation between the GC mass and absolute magnitude in F150W as a function of GC age and metallicity, calculated using PARSEC CMD3.7 single stellar population models \citep{Bressan2012, Marigo2013}. We compute the K-corrected absolute turnover magnitude and use their Equation~5 to find the turnover mass. We estimated age based on the cluster redshift with assumed formation of GCs at a typical epoch $z=4.0$. We fixed the metallicity at $\feh = -0.5$. For simplicity, we use this relationship for both F150W and F200W bands, because the observed F150W--F200W color index is close to zero \citep{2025HarrisReinaCampos, Berkheimer2026, Hinrichs2026}.
Then we estimate $\beta_{\rm eff}$ at $\log (M/\Msun) = 6.0$, in the middle of the range where we fit the power law, so that we can compare the observed slope of the MF with our results.

From the \citealt[Equation~5]{2024Harris} relation we can also estimate the limiting GC mass corresponding to the limiting absolute magnitude of 50\% completeness, which is reported in Table~\ref{tab: obs_limits}. The mass-to-light ($M/L$) ratio in NIR filters tends to be lower than in optical filters. It also decreases with increasing redshift as GCs get younger. We find the range 0.57--0.79~$\Msun/\text{L}_\odot$ for the clusters in Table~\ref{tab: obs_limits}.

\cite{2024Harris} fit the LF for Abell~2744 at $z = 0.308$ and found a turnover magnitude $m_0 = 31.66$ and a dispersion $\sigma_m = 1.39$. This corresponds to $\beta_{\rm eff} \approx 2.0$. At $z = 0.3$ we measure $\beta = 1.87 \pm0.03$, $\beta = 1.95 \pm0.04$, and $\beta = 2.02 \pm0.04$ for TNG100-0, TNG100-17185, and TNG300-0. Uncertainties are the 68\% confidence interval. These model values are close to the Abell~2744 slope.

\cite{Hinrichs2026} fit the LF of PLCK~G165.7+67.0 at $z = 0.35$ for their color-selected and angular-selected GC populations. The best fit for the color-selected population has a turnover absolute magnitude $M_{\rm 0,F200W} = -10.71$ and a dispersion $\sigma_m = 1.2$. The authors note that this gives an unrealistically high turnover mass $\sim10^6\Msun$. Thus, they suggest that their fit with $M_{\rm 0,F200W} = -7.69$ and $\sigma_m = 2.0$ is more robust. This yields an effective power-law $\beta_{\rm eff} \approx 1.8$. For the angular-selected GCs, the authors again point out that the best-fit gives a turnover magnitude too high. They suggest $M_{\rm 0,F200W} = -8.88$ and $\sigma_m = 1.4$ is more reliable, which gives $\beta_{\rm eff} \approx 2.3$. Both of these slopes are quite uncertain. At $z = 0.4$ we measure $\beta = 1.88 \pm 0.03$, $\beta = 1.96 \pm 0.04$, and $\beta = 2.00 \pm 0.03$ for TNG100-0, TNG100-17185, and TNG300-0, which is sufficiently similar to the GC population of PLCK~G165.7+67.0.

\cite{Berkheimer2026} constructed the LF of MACS~J0416.1-2403 at $z = 0.397$ for sources above the 80\% completeness limit. The authors fit an absolute turnover magnitude $M_{\rm 0,F200W} = -8.92$ and a dispersion $\sigma_m = 1.63$. The corresponding effective power law is $\beta_{\rm eff} \approx 1.8$. Our best-fit power-laws at $z = 0.4$ range from $\beta \approx 1.85$ to $\beta \approx 2.0$. The model power laws are moderately steeper than MACS J0416.1-2403. 

\cite{2025HarrisReinaCampos} fit the LF of MACS~J0417.5-1154 at $z = 0.443$ assuming a turnover magnitude $m_0 = 33.1$. They found that a dispersion $\sigma_m \approx 1.6$ matched the data well. These parameters yield $\beta_{\rm eff} \approx 1.9$. From the model at $z = 0.5$, we found $\beta = 1.91 \pm 0.03$, $\beta = 1.97^{+0.05}_{-0.04}$, and $\beta = 2.07 \pm 0.04$ for TNG100-0, TNG100-17185, and TNG300-0, respectively. The model power laws are all similar to MACS~J0417.5-1154.

If we order the observed clusters in redshift, $\beta_{\rm eff}$ does not consistently decrease as we see in the model. This is not well explained by observational incompleteness at high redshift because Abell 2744 at $z = 0.308$ and MACS J0416.1-2403 at $z = 0.397$ achieve similarly deep observations, but $\beta_{\rm eff}$ is steeper for Abell 2744 ($\beta_{\rm eff}\approx 2.0$ to $\beta_{\rm eff}\approx 1.8$). 

The inferred value of $\beta_{\rm eff}$ at a fixed cluster mass (as we take in this work) is also strongly affected by the estimated turnover magnitude. In the samples where turnover magnitude is below the completeness limit, $m_0$ is correlated with $\sigma_m$ and may be underestimated. Fainter $m_0$ would result in steeper effective slope $\beta_{\rm eff}$ and bring the observations in closer agreement with the theoretical predictions. Indeed, we find that the turnover masses inferred for the observational studies referenced here are generally higher than the expected turnover mass $10^5 \Msun$ (see Figure~\ref{fig: mass function}). They ranged from $\log (M_{\rm TO}/\Msun)=5.1-5.3$, except for the color-selected population of PLCK G165.7+67.0 with $\log (M_{\rm TO}/\Msun)=4.8$ where the authors enforced a lower turnover mass. Fitting a power law instead of a log-normal function to the observational samples would allow a more direct comparison with the model results.

\subsection{Surface Density Profile}
\label{sec:SDP_discussion}

We demonstrated that the model surface density profile is steeper at earlier epochs $z \ge0.5$ and thereafter the power law slope is scattered between 1.0 and 1.7. We expect that over time the inner surface density decreases as GCs near the galaxy center are tidally disrupted, and the outer surface density increases as ex-situ GCs are accreted from satellite galaxies. The combined effect is a flattening of the surface density profile. It is possible that this is what we observe at $z \ge0.5$. After $z=0.4$ any underlying trend is obscured by the large scatter.

\cite{Keatley2025} measured the GC surface density profile for the galaxy cluster RXJ~2129 at $z=0.234$ and fit a power law in the range 10--100~kpc. They found the slope $\gamma =1.58 \pm 0.04$ for all sources, $\gamma = 1.92 \pm 0.04$ for metal-rich sources, and $\gamma = 1.42 \pm 0.1$ for metal-poor sources. They split metal-rich and metal-poor sources by color using the threshold F115W--F200W$=0.018$ which corresponds to $\feh = -0.34 \pm 0.2$ based on Equation~4 of \cite{2024Harris}. This is similar to the metallicity threshold used in this work (although this NIR color is not a good proxy for metallicity). Their observations can only probe the mass function down to $\log(M/\Msun) \approx 6.0$. For comparison with \cite{Keatley2025}, we fit a power law to the surface density profile at $z = 0.2$ with $\log(M_{\rm lim}/\Msun) = 6.0$ for all sources and for metal-rich and metal-poor sources using our metallicity threshold $\feh = -0.55$. For all sources, we found $\gamma = 1.75\pm 0.03$, $\gamma = 2.0\pm 0.05$, and $\gamma = 1.50\pm 0.04$ for TNG100-0, TNG100-17185, and TNG300-0, respectively. Here, uncertainties are one standard deviation estimated from a traditional bootstrap analysis with 1000 realizations. For the metal-rich component, we found $\gamma =1.8 \pm0.04$, $\gamma = 2.11 \pm 0.07$, and $\gamma = 1.71 \pm 0.06$ for the three halos. For the metal-poor component, we found $\gamma =1.66 \pm 0.05$, $\gamma = 1.92 \pm 0.07$, and $\gamma = 1.38 \pm 0.05$. The observed power law slopes in \cite{Keatley2025} are within the range of slopes predicted by our model. Across all population, TNG300 appears to be the best match.

\cite{2025HarrisReinaCampos} fit a power law to the surface density profile  MACS~J0417.5-1154 at $z=0.443$ between 10--290~kpc. They found a power law slope $\gamma =1.60 \pm 0.06$ for all sources, $\gamma = 1.68 \pm 0.09$ for red sources, and $\gamma = 1.47 \pm 0.05$ for blue sources. The authors divided their sample in half to distinguish red and blue sources because of a large spread in the F115W--F200W color. We do not attempt to recreate this division. When we apply $\log(M_{\rm lim}/\Msun) = 6.5$ and fit the surface density profile at $z = 0.5$ in the range 10--290~kpc, we find $\gamma = 2.36\pm0.08$, $\gamma = 2.20\pm0.08$, and $\gamma = 1.58\pm0.07$ for TNG100-0, TNG100-17185, and TNG300-0. The slope for TNG300-0 is consistent with the measured value for MACS~J0417.5-1154 within uncertainties.


\cite{Harris2017b} presented the GC surface density profiles of 4 BCGs at $z \approx 0$, and fit a power law in approximately the range 10--100~kpc. They split the profiles into \textit{extreme metal-poor} (EMP) and \textit{extreme metal-rich} (EMR) components to minimize the overlap of the metal-poor and metal-rich modes determined by Gaussian Mixture Modeling (GMM). They defined EMR clusters as GCs with metallicities above the group mean for the metal-rich mode, and EMP clusters as GCs with metallicities below the group mean from the metal-poor mode. The range of slopes they measured for the EMP and EMR components were: $\gamma_{\rm EMP}\approx$ 0.7--1.4 and $\gamma_{\rm EMR}\approx$1.4--2.1. The EMP profile of one galaxy (UGC 10143) was shallow and only modeled with a S\'ersic profile of index $n = 1.5$. We replicated their procedure for dividing the distribution into EMR and EMP components using the \textsc{Scikit-learn} implementation \texttt{GaussianMixture} with two modes to determine the group means. We found a metal-rich group mean $\feh = -0.244$ and a metal-poor group mean $\feh = -1.519$, which we used to define the EMP and EMR populations. At $z = 0.0$ we found a range $\gamma_{\rm EMP}\approx$0.3--0.9 and $\gamma_{\rm EMP}\approx$1.1--1.4.
These values are similar to the values reported by \cite{Harris2017b} though consistently shallower. 
TNG300-0 has a particularly shallow EMP power law $\gamma_{\rm EMP} = 0.34\pm 0.09$ like UGC 10143 which was modeled only with a S\'ersic profile.  

There is considerable variation in the radial profiles of our model halos. Within the range of variation, our model is generally consistent with observed radial profiles. TNG300-0, in particular, was the best match to high redshift clusters. TNG300-0 is the most massive model realization, which could explain the better agreement with similarly massive observed clusters.

\subsection{Half-Mass Radius}

The half-mass radius of the GC population is thought to scale linearly with the mass of the host galaxy. We estimated the expected GC half-mass radius of the giant galaxies studied in this work using the linear relation in \cite{ChenGnedin2022}. This equation estimates the effective radius of the de Vaucouleurs profile, which is conceptually equivalent to the half-mass radius if the surface density profile follows a de Vaucouleurs profile. This relationship was fit using observations of GC populations in Milky-Way-like galaxies. Extending this relationship to massive galaxies, we expect $R_{\rm e} \approx 230$~kpc for the TNG100 model realizations and $R_{\rm e} \approx 550$~kpc for TNG300-0. Our actual measurements of the half-mass radius are much smaller. Averaging across epochs, $R_{\rm e} = 65 \pm 12$ kpc, $R_{\rm e} = 68 \pm 30$ kpc, and $R_{\rm e} = 210 \pm 90$ kpc for TNG100-0, TNG100-17185, and TNG300-0. The reported uncertainty is the standard deviation across epochs. 
These results suggest that the linear relationship in \cite{ChenGnedin2022} does not extend to giant BCG systems. 

\cite{Keatley2025} reported the effective radius of the de Vaucouleurs profile for RXJ~2129 to be $R_{\rm e} = 157 \pm 13$~kpc. This falls between TNG100 $R_{\rm e} \approx 50$~kpc and TNG300 $R \approx 290$~kpc at $z = 0.2$. The mass of RXJ 2129 from weak lensing is $M_{200c} = (7.4 \pm 2.4) \times 10^{14} \Msun$ \citep{Caminha2019}, which is between the masses of the TNG100 halos and TNG300-0, and marginally closer the TNG100 halo mass. The effective radius of RXJ~2129 is consistent with the model, though not tightly constrained given that the masses and effective radii of the model realizations differ significantly.


\subsection{Metallicity}

The metallicity distribution follows a consistent pattern across all three model realizations. There is a prominent high metallicity peak between $\rm{[Fe/H]} = -0.5$ and 0, and a weaker low metallicity peak around $\rm{[Fe/H]} = -1.5$. This bimodal metallicity distribution has long been understood as characteristic of GC populations. We expect that the metallicity distribution evolves at $z<1$ primarily through tidal disruption. Thus, the normalization of the distribution decreases over time. This behavior is observed in the TNG100 model realizations, but not in the TNG300-0 model realization. TNG300-0 gains a significant ex-situ population from a massive satellite galaxy at different epochs depending on whether the satellite is in the 1~Mpc region. This affects the apparent evolution of the metallicity distribution.

We investigated the in-situ and ex-situ cumulative metallicity distributions as a function of the mass threshold and time. As a function of time, the in-situ cumulative distribution profiles decrease monotonically through disruption, while the ex-situ cumulative distribution profiles are strongly influenced by accretion from massive satellites. In both cases, the shape of the profile is consistent. The cumulative metallicity distributions show significant dependence on the mass threshold. At higher mass thresholds, both the in-situ and ex-situ CDFs shift to higher metallicities. In-situ clusters tend to be metal-rich, so the shift to high metallicities corresponds to a shift to a higher in-situ fraction. This suggests that observable populations of GCs systems of BCGs will have higher metallicities and in-situ fractions than the full GC populations, which are largely below observational limits. 

The metallicity dependence on the mass threshold arises from the mass-metallicity relation (blue tilt) observed in the high-mass metal-poor GC populations. We fit the mass-metallicity relation (Equation~\ref{eq: mass-metal relation}) predicted by our model at all 6 epochs. We find the slope above $M_{\rm lim} = 10^6 \Msun$ to be fairly consistent over time, $\alpha = 0.34 \pm 0.02$. Considering the in-situ and ex-situ populations separately, we find almost no blue tilt for in-situ clusters ($\alpha \approx -0.08$) but an even stronger blue tilt for ex-situ clusters ($\alpha \approx 0.42$). Likewise, splitting the distribution at metallicity threshold $\feh = -0.55$, we find no blue tilt for metal-rich GCs ($\alpha \approx -0.03$) and a strong tilt for metal-poor clusters ($\alpha \approx 0.37$). The in-situ and metal-rich slopes are consistent with zero within $2\sigma$. This is expected from observations and previous modeling, which show that the blue tilt is only a property of metal-poor GCs. Since all GCs with $\feh < -1.5$ are formed ex-situ (Figure~\ref{fig:metallicity_cdfs}), we expect the mass-metallicity relation to be a feature of ex-situ clusters but not of in-situ clusters. 

\cite{Choksi2018} generated predictions for the mass-metallicity relation at $z = 0$ using a previous version of the GC model. They found a best-fit value on a stacked sample of model halos of $\alpha \approx 0.23$ with a limiting mass $M_{\rm lim} = 5\times10^5 \Msun$. At $z = 0$ and with a mass threshold of $M_{\rm lim} = 10^6 \Msun$, we measured a steeper slope $\alpha = 0.34$. This deviation is expected because, as noted in \S\ref{sec:metallicity}, there is substantial variation in the blue tilt. \citealt{Choksi2018} found that over many individual model realizations the slope varied from $\alpha = 0$ to $\alpha \approx 0.55$ with a peak around $\alpha \approx 0.3$. 

\cite{Harris2017b} measured the mass-metallicity relation for for metal-poor GCs in 5 BCGs at $z \approx 0$. They were not constrained by strict mass limits as in the high-redshift observations, and fit the relation above the turnover mass $\approx 2 \times 10^5 \Msun$. They found $\alpha \approx 0.25$ for three galaxies, a highly uncertain slope for one, and no trend in the fifth BCG. We replicated their procedure for dividing the distribution into metal-rich and metal-poor components using GMM, and found an average slope over time $\alpha = 0.33\pm0.01$ for the metal-poor GCs above $2 \times 10^5 \Msun$. The observed slope is shallower than our model prediction, but the qualitative trends are similar. 

For high-redshift galaxy clusters observed with JWST, measuring the mass-metallicity relation is challenging. NIRCam color indices such as (F115W--F200W) show a weak metallicity dependence \citep{2024Harris}. Thus the metallicity distribution is difficult to observe, and the mass-metallicity relation cannot be reliably measured. From the modeling we expect that mass-metallicity relation is significant in the GC populations of high-redshift systems, and that the observed samples are biased towards more metal-rich and in-situ formed GCs relative to the complete population.

\subsection{Crossover Radius}

We introduced the crossover radius in the model distributions, where the local population switches from in-situ dominated to ex-situ dominated. We find a clear split between the two regions for TNG300-0, and nearly complete ex-situ dominance for TNG100-17185 (Figure~\ref{fig: in-situ/ex-situ surf den}). For TNG100-0, there is a region between 1 and 50~kpc where the in-situ and ex-situ surface densities are similar. This regions was created by a late merger at $z<1$, which redistributed the in-situ and ex-situ GC populations. TNG100-0 also has the largest initial in-situ fraction of 45\%, significantly higher than the 11\% in TNG100-17185. Both of these factors likely contribute to TNG100-0 and TNG100-17185 samples having significantly different crossover radii despite having similar masses. We conclude that the crossover radius is a sensitive probe of the assembly history of the host galaxy.

We expect that the crossover radius will tend to smaller radii over time as in-situ clusters near the galactic center are disrupted and ex-situ clusters are accreted. This trend is clearly observed for TNG300-0. In TNG100-0, the late merger complicates a simple interpretation of the crossover radius, and we do not necessarily expect it to decrease. In TNG100-17185, the crossover radius is initially very small and there is little change with time. From the present sample of halo realizations, it is not clear if a decreasing crossover radius is a general trend.

We explored how the crossover radius changes with a higher mass threshold, and found that the median change in the crossover radius is less than 1~kpc for both a moderate and significant increase in the mass threshold. At a mass threshold of $\log(M_{\rm lim}/\Msun) = 6.3$ the crossover radius for TNG100-17185 disappears and the system is entirely ex-situ dominated. This is not surprising given that the crossover radius was only $\sim$2~kpc with a mass threshold of $\log(M_{\rm lim}/\Msun) = 5.0$. 

If we use metallicity as a proxy for in-situ and ex-situ clusters, the crossover radius increases by between 10 and 78~kpc (16th-84th percentile range). This occurs because many ex-situ clusters come from giant metal-rich satellite galaxies. We showed in \S\ref{sec:metallicity_classification} that the optimal metallicity threshold still has a $\sim$30\% error (and a $\sim$20\% error in balanced accuracy). These misidentified clusters extend the region of metal-rich dominance and produce a larger crossover radius $R_{\rm cross,m}$. Metallicity, or even a single color, is commonly used in observations to separate in-situ from ex-situ populations. Such measurements of the crossover radius are thus likely to be biased towards higher radii. 

\subsection{In-situ vs. Ex-situ Classification}

We tested two different methods of classifying in-situ and ex-situ GCs based on their observable features: a universal metallicity threshold and machine learning classification. With the metallicity threshold $[\rm{Fe/H}]= -0.55$ evaluated on the full sample, we recovered the correct in-situ and ex-situ populations with a balanced accuracy $A_{B} = 78.5\%$. This metallicity threshold corresponds to the color index F115W--F200W=$-0.03\pm0.04$ \citep[Equation 4]{2024Harris}. Using ML classifiers trained on mass, metallicity, and 2D projected distance we achieved a higher average balanced accuracy $A_{B} = 82.2\%$. For a more practical comparison, we evaluated these classification methods at typical observational completeness limits. With $\log(M_{\rm lim}/\Msun) =6.0$ and 6.3, the optimal metallicity threshold shifts to $[\rm{Fe/H}]= -0.47$ and the balanced accuracy drops to $A_{B} = 74.1\%$ and $71.7\%$. ML classifiers recover slightly higher accuracy $A_{B} = 74.7\%$ and $72.3\%$, respectively.

It is worth noting that these methods were trained and evaluated with different procedures. The metallicity threshold was optimized for each halo realization and redshift separately. A universal metallicity threshold was selected as the average across model realizations and redshifts. The performance of the universal metallicity threshold was evaluated for each model realization and redshift, and the final performance metrics were the average over these evaluations. The ML classifiers were optimized for each redshift and for specific mass and radial thresholds. The three model realizations were combined instead of treated separately. The classifiers were trained on all available observational features and optimized over classifier type and the hyperparameter space. We did not determine a `universal ML classifier'. Each classifier was evaluated on an unseen test dataset at the same redshift that satisfied the same mass and radial thresholds. 

The procedure for the metallicity threshold was designed to create a very simple method for classification, while the procedure for the ML classifiers was designed to create the most accurate classification possible. ML improves the traditional accuracy significantly, by $\sim$12\%. With the different procedures in mind, it is remarkable that on the full sample ML only out preforms the universal metallicity threshold by $\sim$3\% in balanced accuracy. At observational mass limits, the balanced accuracy is nearly equal between the metallicity threshold and ML classifiers. We note that at high mass limits there are far fewer model GCs available to train and evaluate both classification methods. Thus, both classifiers appear to perform worse at observational limits. 

We are primarily concerned with optimizing the balanced accuracy to ensure that the classification truly distinguishes the smaller class (in-situ clusters) from the dominant class (ex-situ clusters). While the metallicity threshold has balanced accuracy error of at least 20\%, adding significantly more complexity does not meaningfully improve the result. This suggests that the metallicity threshold is a good practical choice for observationally classifying in-situ and ex-situ GCs in giant BCGs.

\section{Conclusions} 
\label{sec:conclusion}

We presented predictions of our model for the populations of globular clusters around BCGs of distant galaxy clusters. We investigated evolutionary trends from $z=1$ to $z=0$ in the GC mass function, spatial distribution, mass-metallicity relation, and the overall metallicity distribution. We discussed methods for distinguishing in-situ and ex-situ formed clusters in observations. Here we summarize our main findings. 

\begin{enumerate}

\item The model predicts that the GC mass function becomes shallower as a function of time as low-mass GCs are preferentially disrupted (Figure~\ref{fig: mass function}). Since BCG systems contain massive GCs, which are less affected by disruption, the mean change in the power-law slope is small, from $\beta\approx2$ to $\beta\approx1.9$ (Figure~\ref{fig: mass_function_slopes}). This evolution is even more challenging to detect in current observational samples limited to very massive GCs ($M \gtrsim 10^6\Msun$).
  
\item Outside the BCG core, the model surface density profiles are generally consistent with a power law of projected distance from the galaxy center, but with a wide range of slope $\gamma\approx 1.0-1.8$  (Figure~\ref{fig:SDP_slopes}). The evolution of the slope is balanced by the disruption in inner GCs in strong tidal fields and the accretion of satellite systems (Figure~\ref{fig:surf den}). Of the three model halo realizations we considered, the most massive TNG300-0 provides the closest match to the surface density profiles of the observed high-redshift BCG samples.

\item Accretion of massive satellite galaxies at $z<1$ can dominate the radial profile and metallicity distribution of the overall GC system. In the TNG300-0 halo the largest satellite contributes $\sim$18,000 GCs, many of which are metal-rich (Figure~\ref{fig:metallicity_cdfs}). Major mergers can lead to a significant spatial redistribution of in-situ GCs (Figure~\ref{fig:merger_investigation}).
  
\item Massive GCs follow a mass-metallicity relation above $10^6\Msun$, similar to the blue tilt observed in massive galaxies in the local universe (Figure~\ref{fig:mass-metallicity}). Most of the trend is due to metal-poor ex-situ GCs. In-situ GCs show no tilt. The mass-metallicity relation biases the observed samples in high-redshift systems -- they are expected to be more metal-rich and represent a higher in-situ fraction relative to the full GC population. The strength of the bias depends directly on the observational completeness limits.
  
\item The radial distribution of in-situ and ex-situ GCs is a sensitive probe of the assembly history of the host galaxy (Figure~\ref{fig: in-situ/ex-situ surf den}). The crossover radius between the two populations can vary significantly between host galaxies of similar masses (Figure~\ref{fig:crossoverRscaled}).

\item Machine learning classification performs marginally better than a universal metallicity threshold for classifying in-situ and ex-situ clusters. The additional complexity of ML methods only adds a $\sim$3\% improvement in balanced accuracy (Figure~\ref{fig: balance_accuracy}). Therefore, using a metallicity or color threshold, as is often done in observations, is a justifiable practical choice (Figure~\ref{fig:surf_den_metal}).

\end{enumerate}

More observations of GC systems in high-redshift galaxy clusters are expected from JWST surveys. This investigation provides a theoretical dataset and aids interpretation as the observed samples become more complete.

\begin{acknowledgments}
IS and OYG were supported in part through grants from the Space Telescope Science Institute under NASA contract NAS5-03127 for programs JWST-GO-03433 and JWST-AR-08735.

Claude Opus 5.5 was used to develop and implement our method of determining the metallicity threshold in \S\ref{sec:metallicity_classification}. All outputs were manually inspected and verified. The methodology and code for optimizing machine learning classifiers for in-situ and ex-situ clusters (\S\ref{sec:ML_classification}) was developed with Claude Sonnet 4.6. All output was tested and manually inspected to ensure accuracy. Gemini 4 Hydrogen was used to generate ideas for comparing a log-normal distribution with a power law in \S\ref{sec:mass_func_discussion}. All equations were re-derived by hand to verify accuracy.
\end{acknowledgments}


\facility{JWST}

\software{
NumPy \citep{2020NumPy-Array},
Matplotlib \citep{2007matplotlib},
Scipy \citep{2020SciPy-NMeth},
Scikit-learn \citep{scikit-learn},
Astropy \citep{astropy:2013, astropy:2018, astropy:2022}
Pandas \citep{2010Pandas}
          }



\bibliography{references}{}
\bibliographystyle{aasjournalv7}

\end{document}